\documentclass[a4paper,11pt]{article}

\def\non{\nonumber\\}
\pdfoutput=1 

\usepackage{jheppub} 

\usepackage[T1]{fontenc} 

\definecolor{orange}{rgb}{1,0.5,0}
\definecolor{green}{rgb}{0.2,0.5,0.2}

\title{\boldmath Evidence for  Inflation in an Axion Landscape}

 \author{Pran Nath}
 \author{and Maksim Piskunov}
 \affiliation{Department of Physics, Northeastern University,\\Boston, MA 02115-5000, USA}
\emailAdd{p.nath@northeastern.edu}
\emailAdd{m.piskunov@northeastern.edu}

\abstract{We discuss inflation models  within supersymmetry and supergravity frameworks with a landscape of  chiral 
superfields and one $U(1)$
 shift symmetry which is broken by non-perturbative symmetry breaking terms in the 
 superpotential. We label the pseudo scalar component of the chiral fields axions and their real parts saxions.
 Thus in the models only one combination of axions will be a  pseudo-Nambu-Goldstone-boson which will act as
 the inflaton.
  The proposed models constitute consistent inflation for the following reasons:
 The inflation potential arises dynamically with stabilized saxions,  the axion decay constant can lie
 in the sub-Planckian region, and  consistency with the Planck data is achieved. 
 The axion landscape consisting  of $m$ axion pairs is assumed with the axions in  each pair having opposite charges.
 A fast roll--slow roll splitting mechanism for the axion potential is proposed which is realized with a
 special choice of the axion basis. In this basis the $2m$ coupled equations split into $2m-1$ equations which
 enter in the fast roll and there is one unique  linear combination of the $2m$ fields 
 which controls the slow roll and thus the power spectrum of curvature and tensor perturbations. 
   It is shown that a significant part of the parameter space exists where inflation is successful, 
  i.e., $N_{\rm pivot} = {[50, 60]}$,  the spectral index $n_s$ of curvature perturbations,
  and the ratio $r$ of the power spectrum of tensor perturbations and curvature perturbations,
  lie in the experimentally allowed regions given by the Planck experiment. Further, it is shown that
  the model allows for a significant region of the parameter space where the effective axion decay constant 
  can lie in the sub-Planckian domain. An analysis of the tensor spectral index $n_t$ is also given and the 
   future experimental data which constraints $n_t$ will further narrow down  the parameter space
   of the proposed inflationary models.     
    Topics of further interest include implications of the model for  gravitational waves and  
    non-Gaussianities in the curvature perturbations.  
    Also of interest is embedding of the
    model in strings which are expected to possess a large axionic landscape. 
}

\begin{document} 
\maketitle
\flushbottom

\section{Introduction}
Inflationary models resolve a number of problems associated with Big Bang cosmology which include
the flatness problem,  the horizon problem, and the monopole problem
~\cite{Guth:1980zm,Starobinsky:1980te,Linde:1981mu,Albrecht:1982wi,Sato,Linde:1983gd} 
(for a  review  see~\cite{Linde:2005ht} and for effective field theory of inflation see~\cite{Cheung:2007st}).
In inflation models  quantum fluctuations at the time of horizon exit 
 carry significant information regarding the characteristics of the inflationary model~\cite{Mukhanov+}.
 The cosmic microwave background (CMB) radiation anisotropy  allows  extraction of such characteristics
 which can discriminate  among  models. 
 Thus recently the astrophysical  data  from the Planck experiment ~\cite{Adam:2015rua,Ade:2015lrj,Array:2015xqh} 
 has put  significant constraints on models eliminating many and reducing the parameter space
of others.  One class of  models are those associated with the so called natural inflation
where the inflaton is an axionic field. Thus natural inflation is described by a simple potential~\cite{Freese:1990rb,Adams:1992bn}
\begin{align}
V(a) = \Lambda^4 \left(1+ \cos(\frac{a}{f})\right)\,,
\end{align}
where $a$ is the axion field and $f$ is the axion decay constant.  The Planck data  
requires $f$ significantly greater $M_{Pl}$,  i.e., $f> 10 M_{Pl}$ where $M_{Pl}$ is the Planck mass. 
Now $f>M_{Pl}$ is undesirable since a global symmetry is not preserved  by quantum gravity
unless it has a gauge origin. Further, string theory prefers $f$ below $M_{PL}$~\cite{Banks:2003sx,Svrcek:2006yi}.
 Reduction of the axion decay constant turns out to be a significant problem
 and various procedures have been pursued to overcome it. These include 
 the so called alignment mechanism \cite{Kim:2004rp,Long:2014dta}, the two axion Dante's inferno model~\cite{Berg:2009tg} 
  and  N-flation~\cite{Dimopoulos:2005ac,Easther:2005zr,Grimm:2007hs,Kallosh:2007cc,Olsson:2007he,Battefeld:2007en,Kim:2011jea,Kim:2010ud,Rudelius:2014wla}. 
  Other models using shift symmetry 
  include~\cite{ArkaniHamed:2003mz,Kaplan:2003aj,Green:2009ds,Arvanitaki:2009fg} (for a review 
  and a more extensive set of references  of axionic inflation and of axionic 
  cosmology see~\cite{Pajer:2013fsa,Marsh:2015xka}.). 
  
  In this work we introduce an inflation model in an axion landscape with a $U(1)$ symmetry and 
     with $m$ pairs of chiral fields where the chiral fields in each pair are oppositely charged under 
     the same $U(1)$  global symmetry. 
    We wish to note here that  different authors define the term ``axion'' differently. In the analysis here we will use the 
     term  ``axion''  for  the pseudo-scalar component of 
     any chiral field  and  the corresponding real part will be called a  ``saxion''. 
          In our analysis we have only one
     $U(1)$ global symmetry and thus breaking of it would lead to only one pseudo-Nambu-Goldstone-boson (PNGB)
   and the remaining pseudo-scalars are not PNGBs. It is  important to keep this distinction in mind since sometimes the term
   ``axion'' is automatically interpreted as being a PNGB which is not the case in the analysis here.      
 Returning to the construction of our model the superpotential is chosen to consist of two parts where
     one part is invariant under the $U(1)$ symmetry and the other consists of a symmetry breaking piece
      such as the one arising from instanton effects.  Here we show that 
      a fast roll-slow roll splitting mechanism exits
      which allows a decomposition of the axion potential into a fast roll and a slow roll part where
     inflation is driven by the slow roll part. This set up reduces a multi-field coupled 
      axion system with $2m$ axions      
      to an effective single axion field potential which controls inflation. Using this set up
      we analyze both supersymmetry and supergravity models and show that under the 
      constraints of stabilized saxions, one can find inflation models with the axion 
      decay constant $f<M_{Pl}$ consistent with the data from the Planck experiment. \\
  
    The outline of the rest of the paper is as follows. In section 2, we describe the supersymmetric model 
    of inflation for an axionic landscape consisting of $m$ pairs of axions  with each pair oppositely charged
    under the $U(1)$ symmetry. 
    The superpotential consists of two terms: a part which is invariant 
    under the $U(1)$ symmetry and a part which breaks it arising possibly from instantons.
      Using this superpotential we deduce the conditions for stabilized saxions and then compute the 
      scalar potential for the axions under the constraints of stabilized saxions.  In this section we also
      compute the axion mass matrices.  
            In section 3, we carry out a decomposition of the scalar axion potential
      into a fast roll and a slow part.  To accomplish this we first find the basis where $2m-1$ axions are heavy
      and one axion is light, i.,e., massless  in the limit when there is no breaking of the shift symmetry. 
 Part of the potential which contains the 
heavy fields produces the fast roll inflation while the part that contains the light field generates the 
slow roll part. The decomposition reduces the multi-field inflation to a single field inflation. 
Here we also show that in the slow roll part an effective axion decay constant  $f_e$ enters which is
  given by $f_e=\sqrt {2m} f$ where $f$ is the common decay constant of the 
  axions that enter the superpotential. 
  This result was first derived in  ~\cite{Dimopoulos:2005ac} but here we give a more general derivation of it.
   In section 4, we extend the analysis of section 3 to supergravity with similar conclusions. 
In section 5, we give 
an analysis of the number of e-foldings, of the power spectrum  for curvature and tensor perturbations, 
and of the scalar and of the tensor spectral indices. Specifically we show that much of the allowed parameter space
of experiment  is accessible in this model and
future experiment will constrain the model more stringently. 
In this section we also show how the cosine functions generate  a locally flat potential  necessary for inflation. 
  Conclusions are given in section 6.  In appendices A and B we  define notation and give some mathematical 
  background useful for the analysis carried out in section 5. In Appendix C we  illustrate 
  the emergence of a flat  inflation potential arising from the superposition of cosine functions.

\section{A supersymmetric model of  inflation for an axionic landscape}
  We discuss here a general supersymmetric  framework  for axionic inflation to
   occur\footnote{For references to early work in supersymmetry and supergravity 
see ~\cite{Nath:2016qzm,Ellis:1982ws,SUSYa,Kors:2004hz,SUSYb,BlancoPillado:2004ns,Krippendorf:2009zza,Cicoli:2016olq}.}.
The axions we consider are not QCD axions~\cite{Peccei:1977ur,Weinberg:1977ma,Wilczek:1977pj}
 which were originally the basis of the analysis of \cite{Freese:1990rb}.
In string theory axions occur which are not related to the QCD axions~\cite{Svrcek:2006yi,Halverson:2017deq}.
 Thus we consider  the existence of 
a shift symmetry and assume that in an axionic landscape, such as the one that one might expect in string theory,
 there are a number of axionic fields carrying 
the same $U(1)$ quantum number.  
Now suppose  we have a set of fields $S_i$ ($i=1, \cdots, m$) where $S_i$ carry the same charge under the shift symmetry and the fields
$\bar S_i$ ($i=1, \cdots, m$) carry the opposite charge. Thus under $U(1)$  transformations one
has
\begin{align}
S_i\to e^{i q \lambda} S_i, ~~\bar S_i\to e^{-i q \lambda} \bar S_i, ~~i=1, \cdots, m\,.
\end{align}
The superfields ${S}_{i}$ have an expansion, 

\begin{align}
{S}_{i} = {\phi}_{i} + \theta  {\chi}_{i} + \theta  \theta  {F}_{i}\,,
\end{align}
where ${\phi}_{i}$ is a complex scalar field consisting of the saxion (the real part) and the axion (the imaginary
part), ${\chi}_{i}$ is the axino, and ${F}_{i}$ is an auxiliary field.
Similarly the superfields $\bar {S}_{i}$ have an expansion:
$\bar {S}_{i} = \bar {\phi}_{i} + \bar \theta  \bar {\xi}_{i} + \bar\theta \bar  \theta  \bar{F}_{i}$.
We now consider a superpotential of the form 
\begin{align}
W = W_s(S,\bar S) + W_ {sb} (S, \bar S)\,,
\label{wsn}
\end{align}
where $W_s$ is the part that depends on the fields $S_i,  \bar S_i$ and is invariant under the shift symmetry.
$W_{sb}$ is a  part which breaks the shift symmetry and has the form

\begin{align}
W_{sb} = \sum_{i} A_i(S, \bar S) e^{-T_i}\,,
\end{align}
 where $T_i$ is the action of the $i$-th instanton. In general gauge invariance and holomorphy allow non-perturbative
 terms of the type
 \begin{align}
  A \frac{S^n}{M_{P}^{n-3}} e^{-T}, ~~\bar A \frac{\bar S^n}{M_{P}^{n-3}} e^{-T}\,.
 \end{align}
Detailed structure will depend on the instanton zero modes (see  \cite{Halverson:2017deq,Cvetic:2009ez}
 and the references there in).
Thus we assume the following forms for $W(S, \bar S)$\footnote{A simplified version of this form of the superpotential
has been considered recently in the context of an ultralight axion~\cite{Halverson:2017deq}.}

\begin{equation}
W \left( S,\bar S \right) = \sum_{k = 1}^{m} \sum_{l = 1}^{m} \left( {\mu}_{k  l} {S}_{k} {\bar S}_{l} + \frac{{\lambda}_{k  l}}{2 M} {\left( {S}_{k} {\bar S}_{l} \right)}^{2} \right) + \sum_{k = 1}^{m} \sum_{l = 1}^{q} {A}_{k  l} {S}_{k}^{l} + \sum_{k = 1}^{m} \sum_{l = 1}^{q} {\bar{A}}_{k  l} {\bar{S}}_{k}^{l}\,.
\label{wsn1}
\end{equation}
Here the terms in the first brace on the right hand side are invariant under the shift symmetry  
while the remaining terms on the right hand side violate the shift symmetry.
 The variation of the superpotential
 with respect to $\phi_k$ and $\bar \phi_k$ generate  the constraints that determine the VEVs of $\phi_k$ and 
 $\bar\phi_k$. We assume CP conserving vacua so that the VEVs of the CP odd axionic fields vanish while we set
 $f_k=<\phi_k>$ and $\bar f_k= <\bar \phi_k>$. The constraint equations arising from the variation of the
 superpotential with respect to $\phi_k$ and $\bar \phi_k$ are  
\begin{equation}
\begin{aligned}
\frac{\partial W \left( \phi , \overline{\phi} \right)}{\partial {\phi}_{k}} = 0\,, \\
\frac{\partial W \left( \phi , \overline{\phi} \right)}{\partial {\overline{\phi}}_{k}} = 0\,.
\label{saxion-1}
\end{aligned}
\end{equation}
 We may parametrize $\phi_k$ and $\bar \phi_k$ so that 
\begin{align}
\phi_k = (f_k + \rho_k) e^{ia_k/f_k}, ~~~\bar\phi_k = (\bar f_k + \bar \rho_k) e^{i\bar a_k/\bar f_k}\,,
\end{align}
where $f_k= <\phi_k> ,~\bar f_k= <\bar\phi_k>$ and  $(\rho_k, a_k)$ and $(\bar \rho_k, \bar a_k)$ 
are the fluctuations of the quantum fields around their vacuum expectation values  $f_k(\bar f_k)$. 
They are constrained by the stability conditions for the saxions Eqs. (\ref{saxion-1}) which give 
 \begin{equation}
\begin{aligned}
\sum_{l = 1}^{m} \left( {\mu}_{k  l} {\bar{f}}_{l} + \frac{{\lambda}_{k  l}}{M} {f}_{k} {\bar{f}}_{l}^{2} \right) + \sum_{l = 1}^{q} l  {A}_{k  l} {f}_{k}^{l - 1} = 0\,,\\
\sum_{l = 1}^{m} \left( {\mu}_{l  k} {f}_{l} + \frac{{\lambda}_{l  k}}{M} {f}_{l}^{2} {\bar{f}}_{k} \right) + \sum_{l = 1}^{q} l  {\bar{A}}_{k  l} {\bar{f}}_{k}^{l - 1} = 0\,.
\label{ssb}
\end{aligned}
\end{equation}
We focus here on the scalar potential for the axions and thus we expand around the minimum of the potential
of the saxions, i.e., we
set $\rho_k=0=\bar \rho_k$.
Using the saxion stability conditions given by Eq.(\ref{ssb}) a somewhat lengthy computation gives
for the scalar potential 
 \begin{align}
 \label{v1}
& V(a, \bar a) = V_s(a, \bar a) + V_{sb}(a, \bar a)\,,\\
\non
&{\rm where}\non
\label{v2}
&V_s \left( a , \overline{a} \right)
= \sum_{k = 1}^{m} \big( \sum_{l = 1}^{m} \sum_{p = 1}^{m} \big( {\mu}_{k  l} {\mu}_{k  p} {\overline{f}}_{l} {\overline{f}}_{p} \text{cos} \left( \frac{{\overline{a}}_{p}}{{\overline{f}}_{p}} - \frac{{\overline{a}}_{l}}{{\overline{f}}_{l}} \right) + {\mu}_{l  k} {\mu}_{p  k} {f}_{l} {f}_{p} \text{cos} \left( \frac{{a}_{p}}{{f}_{p}} - \frac{{a}_{l}}{{f}_{l}} \right)\non
&+ \frac{{\lambda}_{k  l}}{M} \frac{{\lambda}_{k  p}}{M} {f}_{k}^{2} {\overline{f}}_{l}^{2} {\overline{f}}_{p}^{2} \text{cos} \left( 2 \frac{{\overline{a}}_{p}}{{\overline{f}}_{p}} - 2 \frac{{\overline{a}}_{l}}{{\overline{f}}_{l}} \right) 
+ \frac{{\lambda}_{l  k}}{M} \frac{{\lambda}_{p  k}}{M} {f}_{l}^{2} {f}_{p}^{2} {\overline{f}}_{k}^{2} \text{cos} \left( 2 \frac{{a}_{p}}{{f}_{p}} - 2 \frac{{a}_{l}}{{f}_{l}} \right) \non
&+ 2 {\mu}_{k  l} \frac{{\lambda}_{k  p}}{M} {f}_{k} {\overline{f}}_{l} {\overline{f}}_{p}^{2} \text{cos} \left( \frac{{a}_{k}}{{f}_{k}} + 2 \frac{{\overline{a}}_{p}}{{\overline{f}}_{p}} - \frac{{\overline{a}}_{l}}{{\overline{f}}_{l}} \right) 
+ 2 {\mu}_{l  k} \frac{{\lambda}_{p  k}}{M} {\overline{f}}_{k} {f}_{l} {f}_{p}^{2} \text{cos} \left( \frac{{\overline{a}}_{k}}{{\overline{f}}_{k}} + 2 \frac{{a}_{p}}{{f}_{p}} - \frac{{a}_{l}}{{f}_{l}} \right) \big)\,, \\ 
&{\rm and }\non
&V_{sb}(a,\bar a) =
 \sum_{l = 1}^{m} \sum_{r = 1}^{q} \big( 2 r  {A}_{k  r} {\mu}_{k  l} {f}_{k}^{r - 1} {\overline{f}}_{l}  \text{cos} \left( \left( r - 1 \right) \frac{{a}_{k}}{{f}_{k}} - \frac{{\overline{a}}_{l}}{{\overline{f}}_{l}} \right)\non
& + 2 r {\overline{A}}_{k  r} {\mu}_{l  k} {\overline{f}}_{k}^{r - 1} {f}_{l} \text{cos} \left( \left( r - 1 \right) \frac{{\overline{a}}_{k}}{{\overline{f}}_{k}} - \frac{{a}_{l}}{{f}_{l}} \right)\non
&+ 2 r  {A}_{k  r} \frac{{\lambda}_{k  l}}{M} {f}_{k}^{r} {\overline{f}}_{l}^{2} \text{cos} \left( \left( r - 2 \right) \frac{{a}_{k}}{{f}_{k}} - 2 \frac{{\overline{a}}_{l}}{{\overline{f}}_{l}} \right) 
+ 2 r {\overline{A}}_{k  r} \frac{{\lambda}_{l  k}}{M} {\overline{f}}_{k}^{r} {f}_{l}^{2} \text{cos} \left( \left( r - 2 \right) \frac{{\overline{a}}_{k}}{{\overline{f}}_{k}} - 2 \frac{{a}_{l}}{{f}_{l}} \right) \big) \non
&+ \sum_{l = 1}^{q} \sum_{r = 1}^{q} \left( l  r  {A}_{k  l} {A}_{k  r} {f}_{k}^{l + r - 2}  \text{cos} \left( \left( r - l \right) \frac{{a}_{k}}{{f}_{k}} \right) + l  r {\overline{A}}_{k  l} {\overline{A}}_{k  r} {\overline{f}}_{k}^{l + r - 2} \text{cos} \left( \left( r - l \right) \frac{{\overline{a}}_{k}}{{\overline{f}}_{k}} \right) \right) 
 \big)\,,
\label{v3}
\end{align}
Here $V_s$ is part of the potential that preserves the shift symmetry and $V_{sb}$ is the part that breaks the 
shift symmetry.  We note that because of the periodic nature of the potential, 
Eqs.(\ref{v1}-\ref{v3}) provide a valid theory even for $a_i > M_{Pl} $.

Next we look at the mass matrix for the axions. The mass matrix consists of three types of terms:
$M \left( {a}_{l} , {a}_{p} \right)$
 which involves only the axions $a_i, i=1, \cdots, m$, $M \left( {\overline{a}}_{l} , {\overline{a}}_{p} \right)$ 
 which involves only the axions $\overline a_i, i=1, \cdots, m$, and the matrix with the cross terms 
 $M \left( {a}_{k} , {\overline{a}}_{p} \right)$.
  For the computation of the heavier axion masses 
 it is sufficient to look at the mass matrix in the limit when shift symmetry breaking  is absent, i.e., ignore $V_{sb}$,  
  and inclusion 
 of the breaking of the shift symmetry would make only negligible contribution to the heavy
 axion masses.  An explicit computation of these gives the following: 
For  $M \left( {a}_{l} , {a}_{p} \right)$ one has 

\begin{equation}
M \left( {a}_{l} , {a}_{p} \right) =2{\delta}_{l  p} \frac{1}{{f}_{p}^{2}} {\left( \sum_{k = 1}^{m} {\mu}_{p  k} {\overline{f}}_{k} \right)}^{2} +2 \sum_{k = 1}^{m} \left( {\mu}_{l  k} + 2 \frac{{\lambda}_{l  k}}{M} {\overline{f}}_{k} {f}_{l} \right) \left( {\mu}_{p  k} + 2 \frac{{\lambda}_{p  k}}{M} {\overline{f}}_{k} {f}_{p} \right)\,.
\end{equation}

For $M \left( {\overline{a}}_{l} , {\overline{a}}_{p} \right)$ one has 

\begin{equation}
M \left( {\overline{a}}_{l} , {\overline{a}}_{p} \right) =2 {\delta}_{l  p} \frac{1}{{\overline{f}}_{p}^{2}} {\left( \sum_{k = 1}^{m} {\mu}_{k  p} {f}_{k} \right)}^{2} + 2\sum_{k = 1}^{m} \left( {\mu}_{k  l} + 2 \frac{{\lambda}_{k  l}}{M} {f}_{k} {\overline{f}}_{l} \right) \left( {\mu}_{k  p} + 2 \frac{{\lambda}_{k  p}}{M} {f}_{k} {\overline{f}}_{p} \right)\,,
\end{equation}

and for the  cross term $M \left( {a}_{k} , {\overline{a}}_{p} \right)$ one has 
\begin{equation}
M \left( {a}_{k} , {\overline{a}}_{p} \right) = M \left( {\overline{a}}_{p} , {a}_{k} \right)
= - 2\sum_{l = 1}^{m} \left( 2 \frac{{\lambda}_{k  p}}{M} \left( {\mu}_{k  l} {\overline{f}}_{l} {\overline{f}}_{p} + {\mu}_{l  p} {f}_{l} {f}_{k} \right) - {\mu}_{k  p}  \left( \frac{{\lambda}_{k  l}}{M} {\overline{f}}_{l}^{2} + \frac{{\lambda}_{l  p}}{M} {f}_{l}^{2} \right) \right)
\end{equation}
This mass matrix is $2m\times 2m$ dimensional. It has $2m-1$ non-zero eigenvalues and one eigenvalue is 
identically zero which corresponds to the inflaton.  
This result is a consequence of the Goldstone theorem\cite{goldstone} which requires that the spontaneous breaking of a single global $U(1)$ 
symmetry leads to a single massless Goldstone mode which implies that there is just one possibility for the inflation field which
ia the zero mode in the diagonalization of a $2m\times 2m$ matrix as noted above and all other modes are heavy after
spontaneous breaking.


\section{The axion landscape and a Fast roll-Slow roll splitting mechanism}

As discussed in the preceding section, even though we have a landscape of axions, i.e., pseudo scalar fields, there is only one $U(1)$ shift symmetry and 
 correspondingly there is only one linear combination of the axion fields which is the pseudo-Nambu-Goldstone-boson and acts
 as the inflation field. The reduction of the multi axion system to a single inflation field does not mean that one has the same
 dynamics if one started with a single field. The reason for this is the follows: our starting point was a landscape of axions 
 each of which undergoes a shift  under the same $U(1)$. The inflation field  thus includes pieces of each of these fields.
 Further,  although in the limit of no breaking of the  shift symmetry, the dynamics of the  (pNGB) inflaton is decoupled from the 
 rest of the axions, there is mixing between the two sectors once the shift symmetry is broken. In this case  the diagonalization 
 will yield   $2m-1$ massive axions and a relatively lighter inflation field.  In principle all the axions both light and heavy 
 enter in inflation, except that the $2m-1$ axions produce a fast roll and die off relatively quickly while the inflation field 
 produces the slow roll. To check that the multi-field analysis is faithfully reproduced by the effective single field, we carry out 
 a numerical analysis for a multi-field case and compare the result to that for the effective single field and find that the 
 fast roll--slow roll decomposition is justified. 
 We emphasize that  the assumption of only one global symmetry will automatically result in only one inflation direction and one inflation candidate and all
 other axions will be heavy.  After the breaking of the shift symmetry, the potential for the axions will in general be mixed involving all the axions.
  We discuss below the explicit procedure for the splitting  of the total potential given by Eqs.(\ref{v1}-\ref{v3}) 
 into the  fast roll-slow roll parts.

The potential of Eqs.(\ref{v1}-\ref{v3}) contains two parts:  $V_s$ and $V_{sb}$ where 
$V_s$ is part of the potential that preserves the shift symmetry and $V_{sb}$ is the part that breaks the 
shift symmetry.  
 Eqs.(\ref{v1}-\ref{v3}) contain a mixture of fast roll and slow roll parts. We wish to decompose 
Eqs.(\ref{v1}-\ref{v3})
to extract the slow roll part. 
There are $2m$ number of axionic fields $a_1, \cdots, a_m$ and $\bar a_1, \cdots, \bar a_m$.
Since there is only one $U(1)$ shift symmetry, 
we can pick a basis where 
only one linear combination of it is variant under the shift symmetry and all others are 
invariant. We label this new basis $b_-, b_+, b_1, b_2, \cdots, b_{m-1}, \bar b_1, \bar b_2, \cdots, \bar  b_{m-1}$ 
where only $b_-$ is sensitive to the shift symmetry. An explicit exhibition of this basis is below 

\begin{align}
{b}_{k} & = \frac{{a}_{k + 1}}{{f}_{k + 1}} - \frac{{a}_{1}}{{f}_{1}},  ~~k=1, 2, \cdots, m-1\,,\non 
{\overline{b}}_{k} &= \frac{{\overline{a}}_{k + 1}}{{\overline{f}}_{k + 1}} - \frac{{\overline{a}}_{1}}{{\overline{f}}_{1}},
~~k=1, 2, \cdots, m-1\,,\non
{b}_{+} &= \frac{{a}_{1}}{{f}_{1}} + \frac{{\overline{a}}_{1}}{{\overline{f}}_{1}}\,,\non
{b}_{-} &= \frac{1}{\sqrt{\sum_{k = 1}^{m} {f}_{k}^{2} + \sum_{k = 1}^{m} {\overline{f}}_{k}^{2}}} \left( \sum_{k = 1}^{m} {f}_{k} {a}_{k} - \sum_{k = 1}^{m} {\overline{f}}_{k} {\overline{a}}_{k} \right)\,.
\label{combinations}
\end{align}
Thus the first three equations in Eq.(\ref{combinations}) give us  $2(m-1)+1= 2m-1$ linear combinations of 
axionic fields which are invariant under the shift symmetry while the last one gives us the combination
of axionic fields which is sensitive to shift symmetry. It can be easily checked that $b_-$ is orthogonal 
to all the rest, i.e., 
\begin{equation}
\left( {b}_{-} , {b}_{k} \right) = 0 =\left( {b}_{-} , {\overline{b}}_{k} \right),
~\left( {b}_{-} , {b}_{+} \right)  = 0\,, ~~~~k=1, 2, \cdots, m-1\,.
\end{equation}
We identify $b_-$ as the inflaton since in the absence of breaking of the shift symmetry it is 
massless while the  remaining  $(2m-1)$ fields $b_+, b_k, \bar b_k, k=1,\cdots, m-1$
are massive. Thus the slow roll is controlled by $b_-$ only. 
Accordingly one can decompose the scalar potential into two parts, $V_{\rm fast}$ and $V_{\rm slow}$ 
where 
\begin{align}
V(b,\bar b)= V_{\rm fast}(b_+, \{b_k\}, \{\bar b_k\}) + V_{\rm slow}(b_-), ~~k=1,\cdots, m-1\,.
\end{align}
 While both $V_{\rm fast}$ and $V_{\rm slow}$ can be computed from Eqs.(\ref{v1}-\ref{v3}), here 
we focus on $V_{\rm slow}$. 
The following projections are useful in extracting the slow roll part of the potential 

\begin{equation}
\begin{aligned}
&\left( {b}_{-} , \left( r - 1 \right) \frac{{a}_{k}}{{f}_{k}} - \frac{{\overline{a}}_{l}}{{\overline{f}}_{l}} \right)  
= \frac{r}{f_e},
&&\left( {b}_{-} , \left( r - 1 \right) \frac{{\overline{a}}_{k}}{{\overline{f}}_{k}} - \frac{{a}_{l}}{{f}_{l}} \right) 
= - \frac{r}{f_e}\,,\\
&\left( {b}_{-} , \left( r - 2 \right) \frac{{a}_{k}}{{f}_{k}} - 2 \frac{{\overline{a}}_{l}}{{\overline{f}}_{l}} \right) =
\frac{r}{f_e},
&&\left( {b}_{-} , \left( r - 2 \right) \frac{{\overline{a}}_{k}}{{\overline{f}}_{k}} - 2 \frac{{a}_{l}}{{f}_{l}} \right) = 
- \frac{r}{f_e}\,,\\
&\left( {b}_{-} , \left( r - l \right) \frac{{a}_{k}}{{f}_{k}} \right) = \frac{r - l}{f_e},
&&\left( {b}_{-} , \left( r - l \right) \frac{{\overline{a}}_{k}}{{\overline{f}}_{k}} \right) = \frac{- r + l}{f_e}\,,
\end{aligned}
\label{projections}
\end{equation}
where 
\begin{align}
f_e= \sqrt{\sum_{k = 1}^{m} {f}_{k}^{2} + \sum_{k = 1}^{m} {\overline{f}}_{k}^{2} }\,.
\label{fe}
\end{align}
Using Eq.(\ref{projections}) in Eq.(\ref{v3}) , and using Eq.(\ref{ssb}) to eliminate $\lambda$, one finds  
\begin{align}
{V}_{\text{slow}} (b_-) =& \sum_{k = 1}^{m} \Big[ 2 \sum_{r = 1}^{q} r  \left( {A}_{k  r} {f}_{k}^{r - 1}  \sum_{l = 1}^{q} l  {A}_{k  l} {f}_{k}^{l - 1} + {\overline{A}}_{k  r}  {\overline{f}}_{k}^{r - 1} \sum_{l = 1}^{q} l  {\overline{A}}_{k  l} {\overline{f}}_{k}^{l - 1} \right) \times \non
& \left( 1 - \text{cos} \left( \frac{r}{ f_e}
{b}_{-}\right) \right)
- 2 \sum_{l = 1}^{q} \sum_{r = l + 1}^{q} l  r  \left( {A}_{k  l} {A}_{k  r} {f}_{k}^{l + r - 2} + {\overline{A}}_{k  l} {\overline{A}}_{k  r} {\overline{f}}_{k}^{l + r - 2} \right)\non
&\times \left( 1 - \text{cos} \left( \frac{r - l}{f_e } {b}_{-}\right) \right) \Big]\,.
\label{slow1}
\end{align}

A remarkable aspect of Eq.(\ref{slow1}) is that the cosine functions depend only on an effective decay constant $f_{e}$. 
Thus for the $2m$ number of fields,  if we set $f_k=f=\bar f_k$,  we have $f_e= \sqrt{(2m)} f$
which means that even with $f$ sub-Planckian we can get the effective $f_{e}$ much larger than 
$M_{Pl}$ if we choose $m$ large enough. 
We note that if we set $f_k= f=\bar f_k$, and $N=2m$, Eq.(\ref{fe}) takes the form $f_e=\sqrt N f$ which is
similar to what one has in the case of N-flation~\cite{Dimopoulos:2005ac}. However, the inflation potential
arising from the fast roll-slow roll splitting
  is very different from the one in N-flation. 
The implication of Eq. (\ref{fe}) and its simplified version  $f_e=\sqrt N f$ is the following:
  by  inclusion of  more fields the range of the axion decay constant
consistent with inflation is enlarged which increases the allowed parameter space of the theory. 
Specifically the region of the sub-Planckian domain of the axion decay constant is significantly 
enlarged.  
One  generally expects this result in the reduction of the type described above,  see, e.g.,  Eq.(3.18) of
\cite{Ernst:2018bib}.
  
\section{\bf Extension to supergravity} 
Next we extend  the analysis  to supergravity where the scalar potential has the form~\cite{Chamseddine:1982jx,Cremmer:1982en}      
  \begin{align}
V= e^K [ D_i W K^{-1}_{ij^*} D_{j^*} W^* - 3 |W|^2] + V_D\,,
\end{align}
Here $K$ is the K\"ahler potential and $W$, as in global supersymmetry, is the superpotential.
Further, 
\begin{align}
D_i W= \frac{\partial W}{\partial \phi_i} + \frac{\partial K}{\partial \phi_i} W\,.
\end{align}
The $D$-term of the potential, $V_D$, will play no role in our analysis and we omit it from here on.
Now in supergravity there  is the well known  $\eta$ problem (see Appendix B for the definition of $\eta$)
 which arises because  the K\"ahler potential contributes to $\eta$ and this contribution can be $O(1)$ while
 for slow roll inflation one needs $\eta<<1$.   To avoid the $\eta$ problem 
we will use the following form for  the K\"ahler potential, i.e, 
\begin{align}
K = \sum_{i} \frac{1}{2} (S_i+ S_i^\dagger)^2\,,
\end{align}
We consider here for simplicity  a single pair of axions with opposite shift symmetries. 
We parametrize  the complex scalar component
$\phi_i$ of $S_i, i=1,2$  so that  
\begin{align}
\phi_i = (\rho_i + i a_i)/\sqrt 2,~ i=1,2\,,
\end{align}
where $ a_i$ have the shift property 
\begin{align}
a_1\to a_1 + \lambda,  ~~a_2\to a_2 - \lambda\,,
\end{align}
and $\rho_i$ are the saxion fields. In this case it is easily checked that the 
kinetic energy is given by
\begin{align}
L_{kin} = - \frac{1}{2} \left[\partial_\mu \rho_i  \partial^\mu \rho_i
+  \partial_\mu a_i \partial^\mu a_i\right]\,,
\end{align}
so the fields $a_i$ and $\rho_i$ are canonically normalized. 
The superpotential is chosen to be of the form 
\begin{align}
W&= W_s + W_{sb}\,,\nonumber\\
W_s& =\frac{1}{2} \mu (S_1+ S_2)^2 +\frac{1}{3} \lambda (S_1+S_2)^3 + W_0\,,\nonumber\\
W_{sb} &= P(S_1)+ P(S_2)\,,
\end{align}
where $W_0$ arises from a hidden sector.
For supergravity analysis the  saxion can be stabilized by imposition of  spontaneous symmetry breaking 
conditions~\cite{Nath:1983aw}
$D_iW =0, i=1,2$ which give 
  \begin{align}
\sqrt 2 \mu f+ 2\lambda f^2  + P'\left(\frac{f}{\sqrt 2}\right) + \sqrt 2f \alpha=0
\end{align}
	where 
	\begin{align}
	 \alpha &=  \mu f^2 + \frac{2\sqrt 2 }{3} \lambda  f^3 + 2 P\left(\frac{f}{\sqrt 2}\right) + W_0\,,
	  \end{align}
	while the vanishing of the vacuum energy after the inflationary period gives the additional constraint
$\alpha=0$. For shift symmetry breaking  we assume 
	 $P= \sum_n A_n e^{c_nS}$ so that 
\begin{align}
W_{sb} = \sum_{n=1}^q A_n \left ( e^{c_nS_1} + e^{c_nS_2}\right)\,.
\end{align}
Next we choose a new basis for the axions, where we replace $a_1, a_2$ by $b_+,b_-$ so that 
\begin{align}
b_{\pm}= \frac{1}{\sqrt 2} (a_1\pm a_2)\,.
\end{align}
where $b_+$ is invariant under the shift symmetry while $b_-$ is sensitive to the shift symmetry. 
As in the global supersymmetry case the 
 computation of the inflation potential is carried out by expanding around the minimum of the saxion potential. Further, we  retain only the slow roll part of the potential which involves $b_-$. In this case $W_{sb}$ takes the form 
\begin{align}
W_{sb}= -\sum_{n=1}^q B_n (e^{i\gamma_n\frac{b_-}{\sqrt 2 f}} + e^{-i\gamma_n \frac{b_-}{\sqrt 2f}})\,.
\end{align}
where $B_n= - A_n e^{c_n f/\sqrt 2}$ and $\gamma_n = c_n f/\sqrt 2$.
In this case,  after stabilization of the saxions and imposition of a vanishing vacuum energy 
at the end of inflation, and using the  fast roll-slow roll splitting mechanism
 we get the following potential for the inflation field 

\begin{align}
V(b_-)&= 
 e^{2f^2}\Big[ 
2\sum_{n=1}^q\sum_{m=1}^q  c_nc_mB_nB_m \left(1- 2 \cos(\frac{\gamma_n b_-}{\sqrt 2 f}) + \cos((\gamma_n-\gamma_m) \frac{b_-}{\sqrt 2 f})\right)\nonumber\\
&+ \sum_{n=1}^q\sum_{m=1}^q 
(16 f^2  + 8 \sqrt 2 f c_n -12)B_nB_m
\Big(1-  \cos(\gamma_n b_-/ \sqrt 2f)-  \cos(\gamma_m b_-/ \sqrt 2f) \nonumber\\
&+\frac{1}{2} \cos((\gamma_n-\gamma_m) b_-/ \sqrt 2f)+ \frac{1}{2}  \cos((\gamma_n+\gamma_m) b_-/ \sqrt 2f)
\Big)
\Big]\,.
\label{sugrapot}
\end{align}
 As noted in the beginning, the 
analysis above is for one pair of axionic chiral fields. As in the global supersymmetry case the analysis  can be
extended to $m$ pairs of axionic chiral fields. Again as 
 for  the global supersymmetry case this extension would modify
the effective axion decay constant for the inflation field so that $\sqrt 2f$ is replaced by $\sqrt{2 m} f$.

\section{Consistency with Planck data~\cite{Adam:2015rua,Ade:2015lrj}}
{\subsection{Experimental observables}}
The path for testing inflationary models with experimental data from the cosmic microwave 
background (CMB) radiation anisotropies has been well laid out in the literature. 
The central quantities that enter are the correlation functions involving scalar and 
tensor perturbations of the  inflaton and of the gravitational field which are 
coupled. From the correlation functions one deduces the  power spectrum for curvature perturbations 
and the power spectrum  for tensor perturbations and the spectral indices. These are the quantities which
are experimentally measured.  A brief description of these is given in section \ref{appendix}
which also
defines the notation we use in this section, i.e., 
${\cal P_R}$ for the power spectrum for comoving curvature perturbations, ${\cal P}_t$ for the  power spectrum
for tensor perturbaions,  and 
$n_s$, $n_t$  for the scalar and tensor spectral indices. 
Inflationary models typically exhibit the power law behavior ${\cal P_R}\propto k^{n_s -1}$ and 
${\cal P}_t\propto k^{n_t}$.
The power spectrum for the curvature perturbations  can be expanded around the pivot scale $k_0$,
where the pivot scale is the  scale  chosen for carrying out the analysis, so that 
\begin{align}
{\cal P_R}(k) & = {\cal P_R}(k_0)  (\frac{k}{k_0})^{n_s(k) -1 }\,,\non
n_s(k) &= n_s(k_0) + \frac{1}{2} (\frac{dn_s }{dln k}) ln(\frac{k}{k_0})\,.
\label{ns}
\end{align}
Similarly for the tensor perturbations we have 

\begin{align}
{\cal P}_t(k) & = {\cal P}_t(k_0)  (\frac{k}{k_0})^{n_t(k)}\,,\non
n_t(k) & = n_t(k_0) + \frac{1}{2} (\frac{dn_t }{dln k}) ln(\frac{k}{k_0})\,.
\label{nt}
\end{align}
One also defines a quantity  $r$ which is the ratio of the tensor to scalar power spectrum so that  
\begin{align}
r=  \frac{ {\cal P}_t(k_0) }{ {\cal P_R}(k_0) }\,,
\label{ratio-r}
\end{align}
The current experimental limits from Planck experiment at $k_0=0.05\, {\rm Mpc}^{-1}$ are ss follows
~\cite{Adam:2015rua,Ade:2015lrj,Array:2015xqh} 
 
\begin{align}
 n_s& = 0.9645 \pm 0.0049\, (68\% {\rm CL})  \non
 r & <0.07\, (95\% {\rm CL})
\label{data}
\end{align}
while $n_t(k_0)$ is currently not constrained.  
As discussed in  Appendix B (sec. (\ref{appendix})) the spectral indices can be related to the slow roll
parameters $\epsilon$ and $\eta$ (see Eq.(\ref{ns-nt-r})).  
Here we will focus on the computation of the spectral indices as these are directly
measured. 
Further, one requires the number of e-foldings to be in the range $N_e=[50,60]$ 
where $N_e$ is the number of e-foldings between the horizon exit and the end of inflation, 
i.e.,  $N_e = N_{\text{total}}-N_{\text{exit}}$, where $N_{\rm exit}$ is the number of e-foldings at 
horizon exit and $N_{\text{total}}$ is the total number of e-foldings at the end of inflation. 
For a mode $k$, horizon exit occurs at  time when $k=R H = \dot R$  and the Hubble radius is $(RH)^{-1}$.
To test our model with Planck data we  use the code of ~\cite{Dias:2015rca} which is suitable for multi field 
inflation models. We have checked this code for simple potentials by a direct integration of the Friedman equations for an inflaton field (see Appendix A). 
 The code provides the spectral indices $n_s,  n_t$ and the ratio of the tensor to scalar
  power spectrum.

\subsection{Monte Carlo Analysis for fit to data}

In this subsection we carry out a Monte Carlo analysis on the parameter space of the models we consider.
First we consider the model of Eq.(\ref{slow1}). However, we will make some simplifying assumptions in the 
analysis which are: ${A}_{k  l} = {\overline{A}}_{k  l} = {B}_{l} {f}^{3 - l}$, ${B}_{l} = B  {G}_{l}$,  ${f}_{k} = f$  for all $k$. Thus $B_l, B, G_l$ are dimensionless while $f_k=f$ carry dimension of mass.
Let us take the potential of Eq.(\ref{slow1}) and simplify it using these assumptions which gives 

\begin{equation}
\begin{aligned}
{V}_{\text{slow}} \left( {b}_{-} \right) = 
4 m  {f}^{4} {B}^{2}  \Big( \sum_{l = 1}^{q} l  {G}_{l}  \sum_{r = 4}^{q} r  {G}_{r}  \left( 1 - \text{cos} \left( \frac{r}{\sqrt{2 m}} \frac{{b}_{-}}{f} \right) \right) \\
- \sum_{l = 1}^{q} \sum_{r = l + 4}^{q} l  r  {G}_{l} {G}_{r}  \left( 1 - \text{cos} \left( \frac{r - l}{\sqrt{2 m}} \frac{{b}_{-}}{f} \right) \right) \Big)\,.
\label{slow2}
\end{aligned}
\end{equation}
Thus Eq.(\ref{slow2}) is the simplified version of Eq.(\ref{slow1}). We remind the reader
 that in  Eq.(\ref{slow2}), $m$ is the number of axion pairs  where 
 the two axions in each pair have opposite  $U(1)$ shift symmetry, and $q$ is the highest power in the polynomial
that breaks the shift symmetry. We assume that the terms in the polynomial 
that break  the shift symmetry are operators with dimension higher than 3 in the 
superpotential. Thus we assume that  $r$ takes on the values in the range $[4,q\geq 4]$. 

Before embarking on a full  Monte Carlo analysis we wish to test the accuracy of the 
fast roll--slow roll decomposition. As shown in sections 3 and 4,
the procedure of decomposing the axion potential into fast roll and  slow roll parts
 brings in a huge simplification  in the
 analysis. We need to verify, however, that neglecting the fast roll part of the potential is a valid approximation.
  For that
 we consider a simple case where we compute spectral indices using the 
 full potential including slow roll and fast roll and then just the slow roll by itself. 
 The case we consider is when we have just one pair of axions, i.e., $m=1$, 
 and we consider a single term in the polynomial that breaks the shift symmetry,
 i.e., we take just the term $G_4$. Further we set $G_4=1$ since it can be 
 absorbed in the parameter $B$. The analysis is done in high precision to exhibit 
 the difference between the two cases. 
 The example we choose is taken from a Monte Carlo analysis and the model point satisfies
 the Planck data constraints. However, $f > M_{Pl}$ in the example below, but this can be easily alleviated by choosing $m > 1$ and the decay constant $f^\prime = f / \sqrt{m}$.
Here we first run the slow-roll part of the potential.
The specific parameters here are
$f = 14.9605 {M}_{Pl}  $, ${b}_{- , 0} = 13.963 {M}_{Pl}  $, and ${N}_{\text{pivot}} = 56$.
 In this case, $n_s = 0.953167 \pm 0.000007$, $n_t = -0.005756 \pm 0.000007$, $r = 0.044045984 \pm 0.000000003$, where errors were computed by varying the number of subhorizon e-foldings in the integration. Next, we run the full (both slow and fast) simulation with $\mu = M_{Pl}$, $B = 10^{-4}$, and $a_1 = \left(b_{-,0} + 13 M_{Pl}\right)/\sqrt{2}$, $\overline{a}_1 = \left(-b_{-,0} - 13 M_{Pl}\right) / \sqrt{2}$. Here one finds that $N_{total}$ increases from $79.2$ to $126.0$ due to the presence of the fast roll part. However, the inflation observables in this case are: $n_s = 0.953169 \pm 0.000006$, $n_t = -0.005754 \pm 0.000006$, and $r = 0.044045988 \pm 0.000000010$. As one can see the accuracy of slow-roll fast-roll decomposition surpasses the precision of the integration.
  We further compare these results to the answers obtained with a slow-roll approximation, from which we get $\epsilon = 0.00288$, $\eta = -0.0150$, 
$r = 0.04604$, 
$n_s = 0.9527$,   $n_t = -0.005755$.
As one can see apart from $n_t$ the discrepancy is now significantly larger than the integration precision. We get $\delta r = 0.002$, $\delta n_s = 0.0005$, $\delta n_t = 0.000001$ comparing to the single-field simulation. 
However, while the $n_t$ value is extremely close to the slow roll prediction, there is a deviation in the value of $r$.  Thus  more precise measurement of $r$,
would allow a test for deviation from the slow roll relation $n_t+r/8=0$.

  Similarly, for a supergravity example on Fig.(\ref{fig7}), we get $r = \left(1.03994375 \pm 0.00000002\right) \times 10^{-4}$, $n_s = 0.9724973 \pm 0.0000007$, $n_t = \left(-1.29 \pm 0.07\right) \times 10^{-5}$. On the other hand, from slow-roll approximation we get, $r = 1.07 \times 10^{-4}$, $n_s = 0.9714$, $n_t = -1.34 \times 10^{-5}$. Here, both values of $r$ are way below experimental limit (as is true for most cases on Fig.(\ref{fig10})), $\delta n_s = 0.0011$, which is $22\%$ of the experimental limit, and the difference for $n_t$ is smaller than the simulations precision.   

\begin{figure}
\centering
\includegraphics[width=1\textwidth]{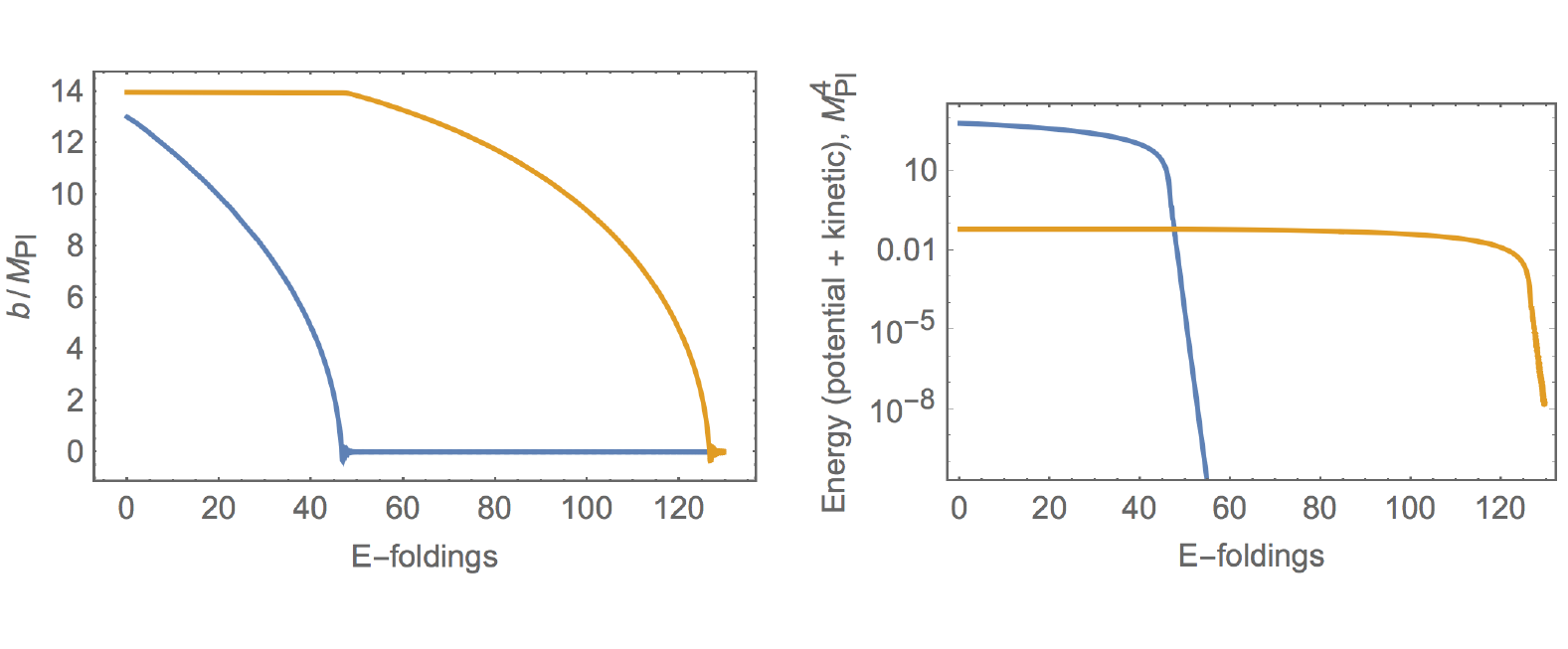}
\caption{Left panel: fast and slow field components as a function of the e-foldings. Right panel: energy of the slow and fast field components as a function of e-foldings. Slow field energy is defined as ${E}_{\text{slow}} = {V}_{\text{slow}} \left( {b}_{-} \right) + \frac{1}{2} {\dot{{b}_{-}}}^{2}$. Fast field energy is defined as ${E}_{\text{fast}} = {V}_{\text{full}} \left( {b}_{-}, {b}_{+} \right) - {V}_{\text{slow}} \left( {b}_{-} \right) + \frac{1}{2} {\dot{{b}_{+}}}^{2}$. One can see that the fast component starts larger, but then the slow component overtakes it. Here $B = {10}^{- 4}$.}
\label{fig1}
\end{figure}

\begin{figure}
\centering
\includegraphics[width=1\textwidth]{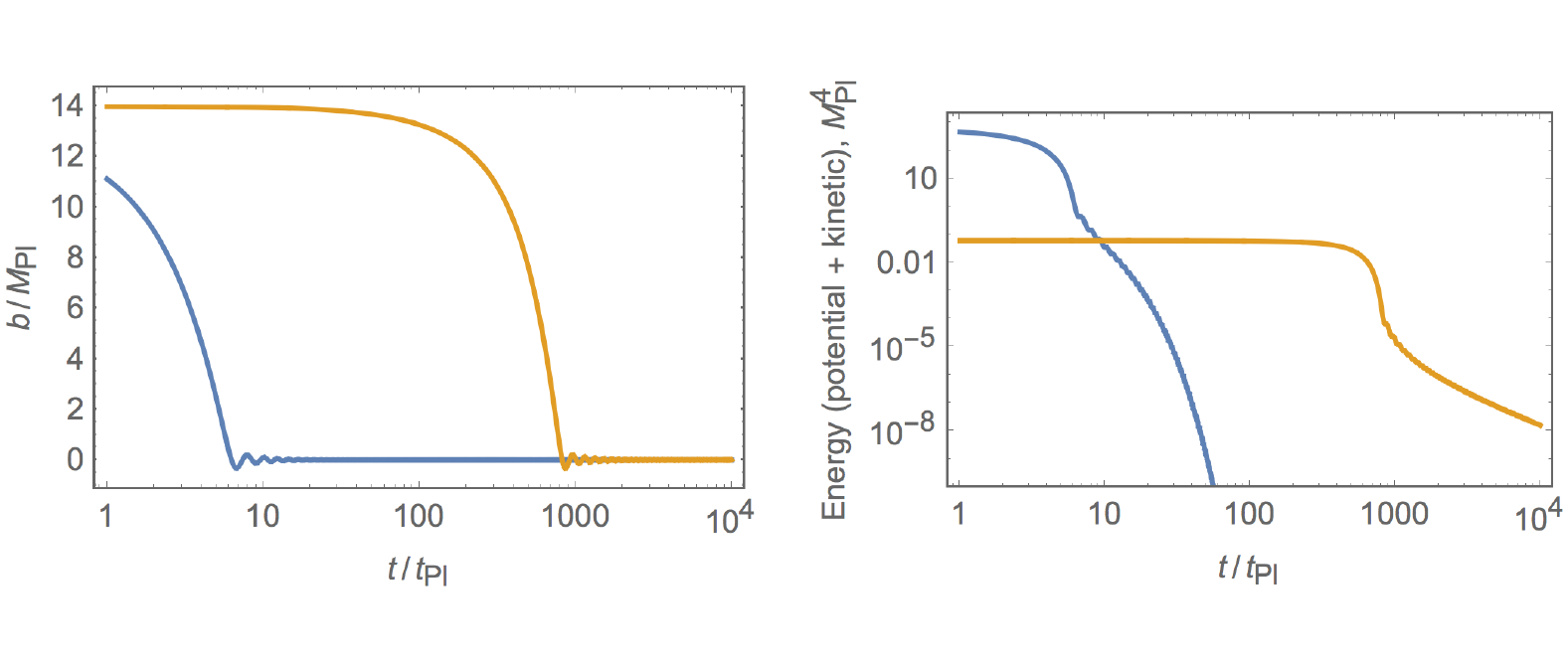}
\caption{Left panel: fast and slow field components as a function of time. Right panel: energy of the slow and fast field components as a function of time. Slow field energy is defined as ${E}_{\text{slow}} = {V}_{\text{slow}} \left( {b}_{-} \right) + \frac{1}{2} {\dot{{b}_{-}}}^{2}$. Fast field energy is defined as ${E}_{\text{fast}} = {V}_{\text{full}} \left( {b}_{-}, {b}_{+} \right) - {V}_{\text{slow}} \left( {b}_{-} \right) + \frac{1}{2} {\dot{{b}_{+}}}^{2}$. One can see that the fast component starts larger, but then the slow component overtakes it. Note the logarithmic scale of time. Here $B = {10}^{- 4}$.}
\label{fig2}
\end{figure}

A comparison of fast roll vs slow roll is given in Fig.(\ref{fig1}). Here the left panel gives the fast and the slow field components as a function of the number of e-foldings. The right panel gives a comparison of the  energy of the slow field and the energy of the fast field components as a function of the number of e-foldings. One finds that the fast component starts with a larger energy, but the slow component overtakes it and drives inflation. In Fig.(\ref{fig2}) we give the  fast and the slow field components as a function of time in the left panel. On the right panel we display the energy of the slow field and of the fast field as a function of time. We note that the fast field component dies about a factor of 100 faster than the slow component in this case. {Similar behavior generally occurs in both global supersymmetry and supergravity models as long as the slow-roll part of the potential is much smaller than the fast-roll part. For that reason we disregard the fast-roll part in the further analysis, and perform simulations with the slow-roll part only.}

\begin{figure}
\centering
\includegraphics[width=1\textwidth]{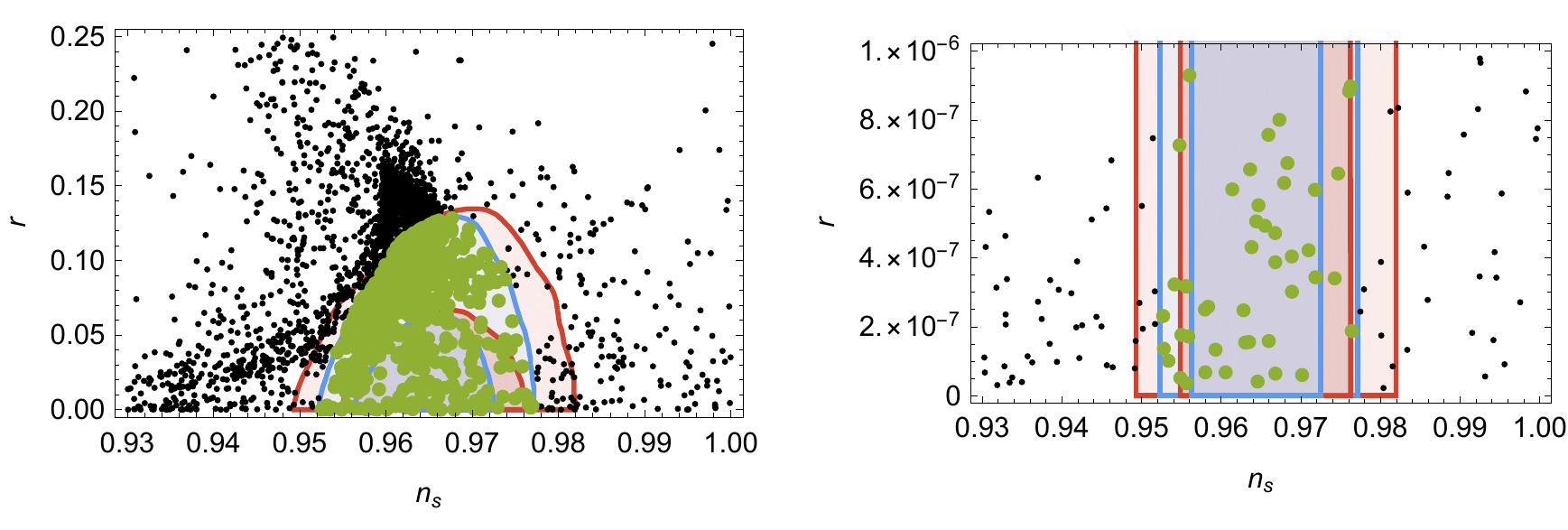}
\caption{
Monte-Carlo analysis for the model with ${G}_{4} = 1$, ${G}_{5} \neq 0$  for the entire input parameters space shown in observables space. The blue region corresponds to $68 \%$  (inner contour) and $95 \%$ CL (outer contour) 
regions of Planck experiment~\cite{Ade:2015lrj}
TT, TE, EE + lowP (include polarization data).
The red regions correspond to Planck TT+lowP (do not include polarization data). For a discussion of 
TT, TE, EE + lowP see reference ~\cite{Ade:2015lrj}. The green and the black scatter points corresponds to scenarios where $b_-$ reaches global minimum at the end of inflation, and the green  scatter points satisfy experimental bounds on $n_s$ and $r$ for the 
TT, TE, EE + lowP  Planck data analysis. All model points have $N_{\rm}=[50,60]$. The decay constant $f < M_{Pl}$ for all points on the right panel.}
\label{fig3}
\end{figure}

We proceed now to discuss the result of the full Monte Carlo analysis of the parameter space of 
Eq.(\ref{slow2}).
In the left panel of Fig.(\ref{fig3}) we give an analysis of $r$ vs $n_s$ for the parameter space when $G_4=1, G_5\neq 0$. The scatter plot is on the Planck data where
the blue region corresponds to 68\% and 95\% CL regions of Planck TT, TE, EE + lowP in Fig. (11) of 
~\cite{Ade:2015lrj}.   One finds that there are 
a large number of parameter points in the region in $r$ vs $n_s$ allowed by the Planck experiment.  
The black dots and the green dots  correspond to scenarios where $b_-$ reaches global minimum at the end of inflation. The scatter points have a mixture of axion decay constant both below and above the 
Planck mass.  In the right panel we exhibit the data set where $f<M_{Pl}$. We note in passing that 
the parameter points in the left panel where $f>M_{Pl}$ may be reduced to lie below $M_{Pl}$ by choosing 
the appropriate number of axion pairs $m$. 

\begin{figure}
\centering
\includegraphics[width=0.75\textwidth]{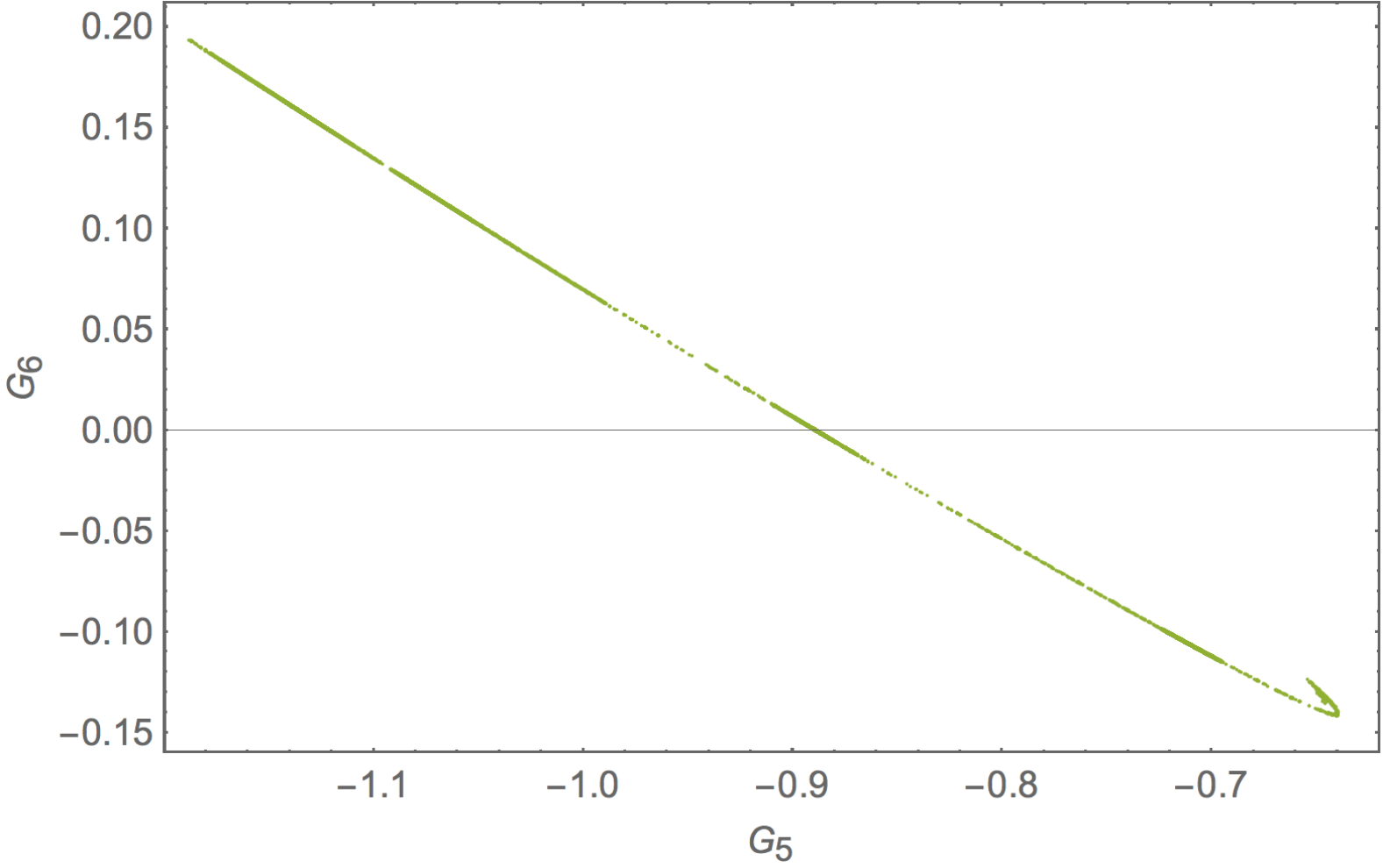}
\caption{
Exhibition of a  region in $\left| {G}_{6} \right| < \left| {G}_{5} \right|$ plane where
all experimental constraints are satisfied. The region exhibited does not exclude other 
regions where consistent inflation can occur. The analysis shows that inflation can occur
in extended regions of the parameter space of models.}
\label{fig4}
\end{figure}

\begin{figure}
\centering
\includegraphics[width=0.8\textwidth]{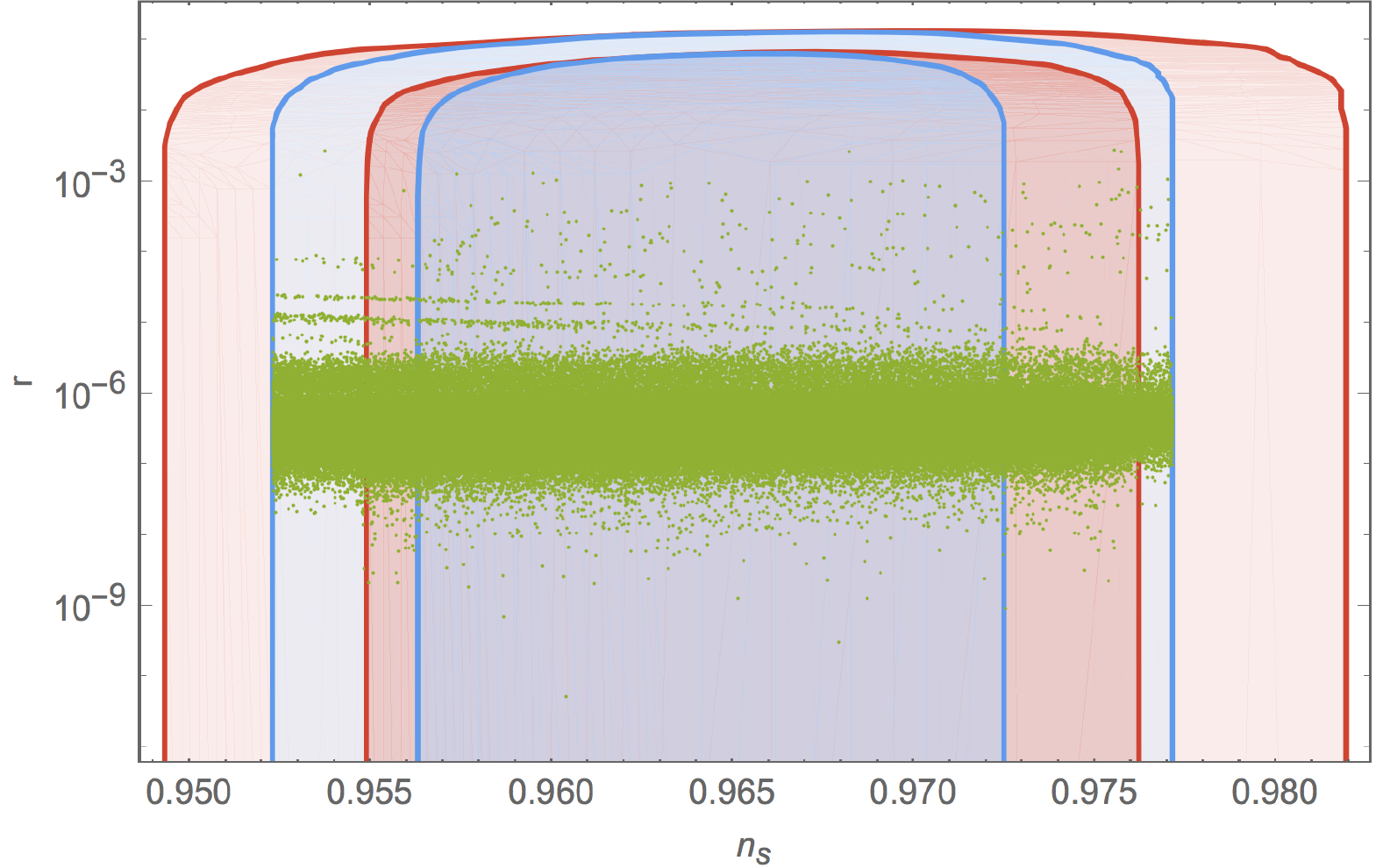}
\caption{
Monte-Carlo analysis when
 ${G}_{4}$ = 1, ${G}_{5}\neq 0$, ${G}_{6}\neq 0$ for the entire input parameters space shown in (${n}_{s}$, $r$) space. 
 The entire set of points shown pass all experimental tests. Note the log scale on the $r$ axis.}
\label{fig5}
\end{figure}

\begin{figure}
\centering
\includegraphics[width=0.8\textwidth]{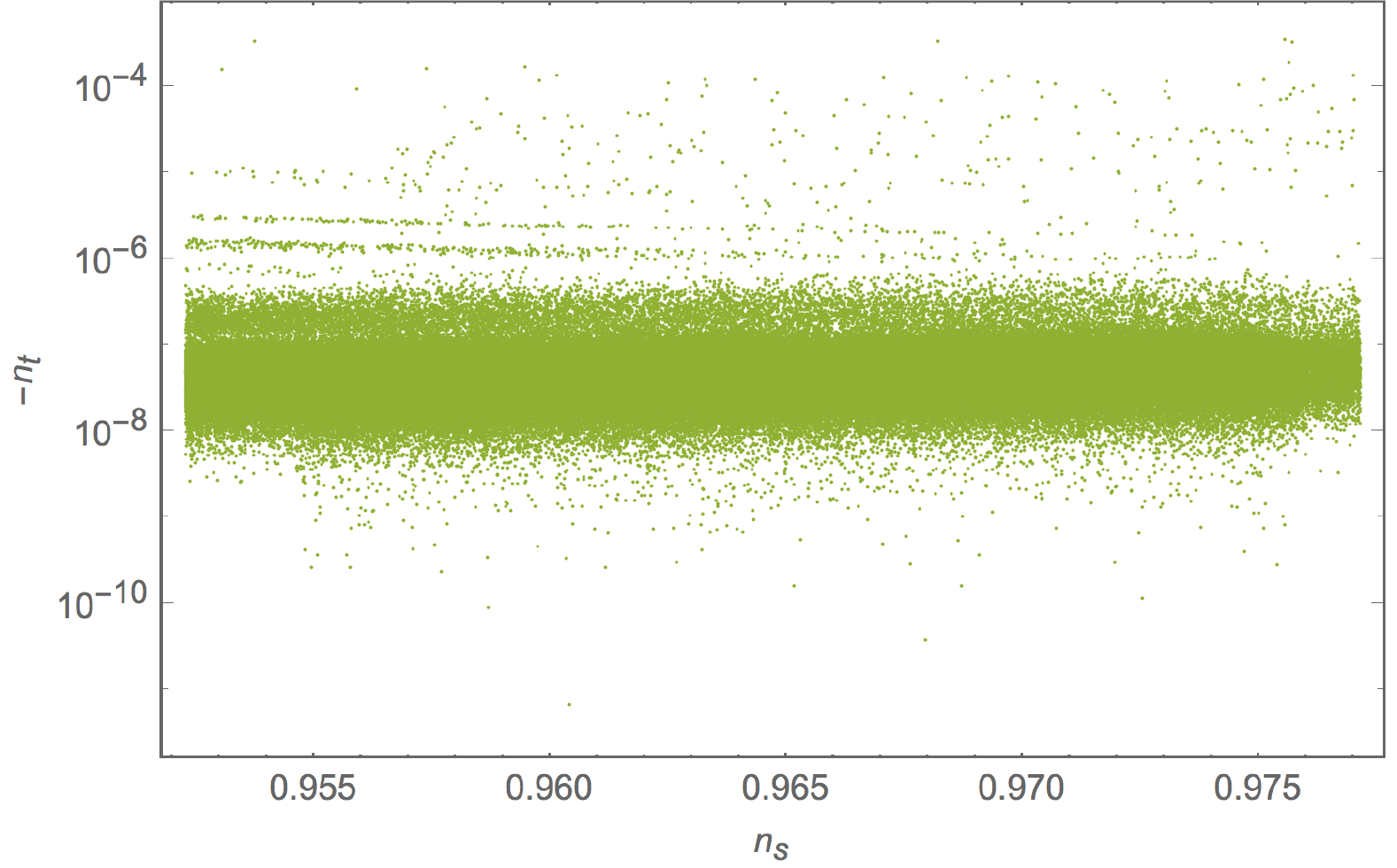}
\caption{
Monte-Carlo analysis when
 ${G}_{4}$ = 1, ${G}_{5}\neq 0$, ${G}_{6}\neq 0$ for the entire input parameters space shown in (${n}_{s}$, $n_t$) space
 for the same  dataset as  in Fig.(\ref{fig5}). Note the log scale on the $-n_t$ axis.}
\label{fig6}
\end{figure}

     Next we  extend our result to include ${G}_{6} \neq 0$. The analysis enlarges the number of allowed models which pass all the experimental tests and have $f<M_{Pl}$. 
   We do this by starting with the parameter points passing experimental tests from the analysis of Fig.(\ref{fig3}) 
  and extending the region corresponding to these points into ${G}_{6} \neq 0$ region.  The analysis shows that
  the inflation trajectory is a narrow strip in the $\left( {G}_{5}, {G}_{6} \right)$ plane as shown in Fig.(\ref{fig4}). 
  In Fig.(\ref{fig5}) we give the distribution in  $\left( {n}_{s} , r \right)$ for the parameter space of Fig.(\ref{fig4}).
We note that here $N_{\rm pivot}= [50, 60]$, the 
 full set of  the parameter points pass the experimental constraints on $n_s$ and $r$ and the parameter points have 
 $f<M_{Pl}$. In Fig.(\ref{fig6}) we give the distribution of $n_t$ vs. $n_s$ for the same parameter space as for Fig.(\ref{fig5}).
 Currently there is no experimental data on $n_t$ and thus the distribution in the ($n_s$, ${n}_{t}$) plane is
 a prediction  which may be tested in future experiments.
Cleary a constraint on $n_t$ will narrow down the range of the parameter space further.  
Here we note that the inflationary trajectory which lies in a narrow strip in the $\left( {G}_{5}, {G}_{6} \right)$  plane 
in Fig.(\ref{fig4})
indicates that inflation with desired properties does not arise for random choices of parameters but occurs along a constrained
path in the $\left( {G}_{5}, {G}_{6} \right)$ plane. Thus inflation  is a fine tuned phenomenon. Part of the reason
for fine tuning is the following: while inflation with a  number of e-foldings can arise for a 
much larger parameter space,  the parameter space  gets 
 successively reduced as one imposes the constraint on  $N_{\rm pivot}=[50,60]$, 
and the constraint on  $n_s$ and on $r$ (see, e.g., Fig.(\ref{fig3})).
In other words 
one does not expect that any random choice of parameters will result in the desired number of e-foldings
and with $n_s$ and $r$ consistent with experiment. Thus inflation occurs in small patches of the parameter
space but the  number of such patches is very large. A similar phenomenon occurs in electroweak symmetry breaking in
supergravity models~\cite{Arnowitt:1992aq}
 where the constraints of color and charge conservation, relic density constraints and the constraint
to have the $Z$-mass to be the experimental value significantly reduce the parameter space of supergravity models.
We note in passing that all the parameter points exhibited in Figs.(\ref{fig3})-(\ref{fig6}) satisfy the 
Lyth bound~\cite{Lyth:1996im}
for slow roll inflation, i.e., 

\begin{align}
\frac{\Delta \phi}{M_{Pl}} \geq \left(\frac{r}{0.01}\right)^{{1}/{2}}\,,
\label{lyth}
\end{align}   
where $\Delta \phi = (\phi_{\rm end} - \phi_{\rm exit})$. We also note in passing that the inflaton mass in the
cases considered above is $O(10^{-4}) M_{Pl}$.  With reference to Fig.(\ref{fig4}) we note that  the purpose of the analysis 
 is to present one concrete example of a trajectory which supports inflation consistent with data 
 but there could be other trajectories which do the same. The generation of such trajectories is computer intensive and a dedicated computer analysis would be needed to exhaust all the allowed regions where inflation can occur.\\

    Finally we give the Monte Carlo analysis for the supergravity model of Eq.(\ref{sugrapot}).
       We begin by displaying in 
    Fig.(\ref{fig7}) a generic potential for $b_-$ 
    for the supergravity case when $m=1$ (see Eq.(\ref{sugrapot})).
    Next for the same parameter set  as used in Fig.(\ref{fig7})   we exhibit the
    phenomenon of fast roll and slow role in Fig.(\ref{fig8}).
Thus in  Fig.(\ref{fig8})
the left panel gives the evolution of the fast component (blue) and of the slow component (red) as a function of  time,
while the right panel gives the energy of the fast (blue) and the slow (red) component as a function of time. 
As in the global supersymmetry case here too one finds that the fast component dies about a hundred times faster than
the slow  component and thus the slow component rules inflation.  
      To test the model with experiment, we note that for the case $q=1$, the desired number of e-foldings are not achieved, while for the case  $q=2$ one can get the right number of e-foldings but not the desired values of $n_s$. However, the case $q=3$ gives  consistent inflation. 
  Here we investigate the range of the parameter space where  $A_1 = M_{Pl}^2$, $A_2/A_1 = [-1, 1]$, $A_3 /A_2 = [-1, 1]$, 
  $f/M_{Pl} = [0, 1]$.  
In Fig.(\ref{fig9}) we exhibit the parameter space which gives rise to consistent inflation. The left panel displays the 
parameter  space in the $A_3/A_1$ vs. $A_2/A_1$ plane and the right panel exhibits the parameters space in the
$f$ vs. $A_2/A_1$ plane. As can be seen in this panel all the parameter space exhibited has $f$ lying in
the sub-Planckian region. 
In the left panel of Fig.(\ref{fig10}) we give the prediction of $r$ vs $n_s$ and  in the right panel  a prediction of
 $n_t$ vs $n_s$. In Figs.(\ref{fig9}) and  (\ref{fig10}) the green and the black points have the same meaning 
  as in Fig.(\ref{fig3}). As in global supersymmetry case, here too all the parameter points exhibited in 
  Figs.(\ref{fig9})-(\ref{fig10}) satisfy the Lyth bound~\cite{Lyth:1996im}.

 The analysis given above shows  that the effective theory  does not automatically lead to the exact standard single-field inflation
   results  but the predictions of the model lie very close to them.  For example, in one of the cases considered above   
   one finds deviations from the standard single-field case so that $\delta r = 0.002$, $\delta n_s = 0.0005$ and 
   $\delta n_t = 0.000001$.  Thus as stated earlier although the $n_t$ value is very close to the slow roll prediction, there is a deviation in the value of $r$ 
   and a  more precise measurement of $r$ would allow  a test for deviation from the slow roll relation $n_t+r/8=0$.

\subsection{Generation of a flat inflation potential}

In this section we give further details on the generation of 
a  locally  flat  scalar potential arising from the instanton induced axion superpotential.
 In  supersymmetric models symmetry breaking terms are added in the superpotential with coefficients 
 which are exponentially suppressed. For $q$ number of symmetry breaking terms induced
 by instantons in the superpotential, there are in general $n=2q$ number of cosines with 
 different periods in the scalar potential which follows from the simple relationship between the 
 scalar potential and the superpotential. Further, the coefficients of the cosines contain products of
 suppression factors because the scalar potential generates cross-terms which automatically arise
 in going from the superpotential to the scalar potential as can be seen from Eq.(38) where a double 
 sum appears on the cosines and their coefficients.  Such cross terms are relevant in generating a
 local flatness of the scalar potential. 
To illustrate this point more clearly we exhibit in
Appendix C (section 9) the analytic expression for the scalar potential for the case $m=1, q=3$ 
relevant for Fig. 7. First, we note that  in this case one has six cosines which are of the form $\cos(\frac{nb}{\sqrt 2 f}), ~n=1,\cdots, 6$
and  we have displayed the suppression factors for each of the six cosines in Eq. (72).
In Eq. (72)  $A_i, i=1,2,3$ are the suppression factors and we see that products of them appear in 
all six cosines. Thus $A_1$ appears in the first four, $A_2$ appears in the first five and $A_3$
appears in the coefficients of all six cosines. In Eq.(73) we give the numerical sizes of the 
coefficients of the cosines for the case 
of Fig. 7. Here we note that for the case when 
$$ \frac{b}{\sqrt 2 f}= \pi$$
(which corresponds to $b= 3.65$ for $f=0.8211$ used in Fig.7)
the terms in the first, the third and the fifth brace in Eq.(73) give maximum contributions 
while the contributions of the terms in the second, the fourth and the sixth brace vanish. 
Thus we see three sets of terms going through a maximum and the other three going through
a minimum. 
 With reference to Fig. 7, we see that indeed the first cosine term is going  through its
 maximum while the second cosine is going through its minimum at $b=3.65$. The higher  cosine terms
 are following similar patterns although their contributions are relatively small.  
A superposition of these leads to a local flatness around the point where the maxima 
and the minima simultaneously occur. 
This can be roughly seen by just superposing the first two 
dominant terms in Eq. (73) 
which appear as the green curve and the brown curve in Fig. 7.
Further from Eq.(73) one can also understand the size of the potential in the 
region of flatness. \\

We note in passing  that a similar procedure of using several non-perturbative terms
to produce inflation by adjustment of parameters in the non-perturbative terms is used in the 
so-called `racetrack' models (see, e.g.,~\cite{BlancoPillado:2004ns,Lalak:2005hr,Greene:2005rn,BlancoPillado:2006he,Sun:2006xv,Brax:2007fe,Wen:2007ek,Brax:2007fz,Wen:2008zz,Brax:2008dn,Chen:2009nk,Badziak:2009eh,Allahverdi:2009rm,Olechowski:2010zz,Badziak:2010qy} and for 
related works~\cite{Higaki:2014pja,Higaki:2014mwa, Kadota:2016jlw,Kobayashi:2016vcx}).
In our analysis the shift symmetry produces a flat direction and its breaking 
 reduces the shift symmetry from a continuous to a discrete one. The breaking  produces a specific 
 combination of cosines with different periods which 
 leads to local flatness in the potential at points where the maxima of one set of cosines 
 overlaps with the minima of the other set of cosines.

\section{Conclusion} 

We have investigated  models of  multi-field inflation when there is an underlying $U(1)$
shift  symmetry.  We have proposed a new technique, the fast roll-slow roll splitting mechanism,
which involves  a decomposition of the inflation potential
into two parts: a fast roll part and a slow roll part.
 The technique was tested and found to be 
highly accurate. The fast roll-slow roll decomposition reduces the problem of multi-field inflation to that of a single field inflation   and is likely to be  realized  in string theory  because of the possible existence of many axions with 
$U(1)$ symmetry in strings. We have applied this technique for the analysis of spectral indices for
an inflation model in a supersymmetric axionic landscape. 
 The breaking of the shift symmetry was accomplished by instanton inspired terms in the superpotential.
       It is found that a large part of the parameter space exists where a number of e-foldings in the range
   $[50,60]$ are accomplished. This parameter space is reduced on imposing the experimental
   limit on the spectral index $n_s$ of curvature perturbations and the limit on the
   ratio of the power spectrum of  tensor perturbations to the power spectrum of curvature perturbations. 
Further, in the analysis we show that inflation, which satisfies all experimental constraints,  
can be achieved with sub-Planckian values of the decay constant which is a significant result of  the
analysis.
 It is found that the axion landscape model can reach 
 most of the allowed region of experiment for the spectral indices.
   The model allows the inflaton field to have initial values which lie in the super-Planckian
region and thus there is a possibility of gravitational waves during the inflationary period. 
Among other topics of interest is the heating after inflation (for a recent review see~\cite{Amin:2014eta})
and a variety of associated phenomena such as non-Gaussianities in the curvature perturbations, 
generation of primordial black holes and of primordial magnetic fields, and baryogenesis.
These are topics worthy of study within the well defined framework presented in this work 
which is a model with stabilized saxions and supports inflation consistent with all of the current 
experimental constraints.   Since strings are the most natural framework where axionic landscape can occur,
it would be natural to extend this analysis to  strings which should be an interesting topic of 
investigation in the future a realistic analyses could be done with stabilized moduli as in 
\cite{kilt,lvs}.

\acknowledgments
Conversations with Jim Halverson, Cody Long  and Fernando Quevedo are acknowledged.
  This research was supported in part by the NSF Grant PHY-1620575.

\section{Appendix A: Preliminaries}
As our starting point we  assume homogeneity on cosmological scales, so that the metric takes the form

\begin{equation*}
ds^2 = - d {t}^{2} + {R\left( t \right)}^{2} \left( d {x}^{2} + d {y}^{2} + d {z}^{2} \right)\,,
\end{equation*}
where $R(t)$ is the scale factor. For $2m$ number of axionic fields, the Lagrangian is given by 

\begin{equation}
\mathcal{L} = - \frac{1}{2}\sum_{a=1}^{2m} \left({\partial}_{\mu} {\phi}_{a} {\partial}^{\mu} {\phi}_{a} \right)
- V \left( \phi \right)\,.
\end{equation}
In this case the two Friedman equations are 

\begin{align}
3 M^2_{Pl} \frac{{\dot{R}}^{2}}{{R}^{2}} = \frac{1}{2} \dot{{\phi}_{a}} \dot{{\phi}_{a}}  + V \left( \phi \right)\,,\non
- 2 M^2_{Pl}\frac{\ddot{R}}{R} - M^2_{Pl}\frac{{\dot{R}}^{2}}{{R}^{2}} = \frac{1}{2} \dot{{\phi}_{a}} \dot{{\phi}_{a}}  - V \left( \phi \right)\,.
\label{F2}
\end{align}

Further the field equations give us 

\begin{equation}
\ddot{{\phi}_{a}} + 3 \frac{\dot{R}}{R} \dot{{\phi}_{a}} = - \frac{\partial V}{\partial {\phi}_{a}}\,.
\label{Fieldeq}
\end{equation}
Let us define the number of e-foldings $N$ so that  $N= ln(R/R_0)$, where $R_0$ is the value of $R$ at the beginning of inflation. 
One  can derive the following equations for 
$N$ from the Friedman equations Eqs.(\ref{F2}) and the field equation Eq.(\ref{Fieldeq})

\begin{align}
3 M^2_{Pl}{\dot{N}}^{2} = \frac{1}{2} \dot{{\phi}_{a}} \dot{{\phi}_{a}} + V \left( \phi \right)\,,\non
\ddot{{\phi}_{a}} + 3 \dot{N} \dot{{\phi}_{a}} = - \frac{\partial V}{\partial {\phi}_{a}}\,.
\label{e-foldings}
\end{align}
Integration of these equations allows for a determination of the axion fields as a function of the  
number of e-foldings  and also as a function of time.

\section{Appendix B: Power spectrum and Spectral Indices\label{appendix}}
To test the inflation model with Planck data ~\cite{Adam:2015rua,Ade:2015lrj}
we need to compute the  power spectrum of curvature
and tensor perturbations and 
spectral indices. A brief discussion is given here to define notation and make the numerical analysis more 
comprehensible, and more exhaustive treatments can be found in several reviews
\cite{Lyth:1998xn,Liddle:2000cg,Langlois:2010xc,Baumann:2009ds}.
 Let us begin by considering a massless scalar field within a gravitational background so that 

\begin{equation}
{\cal A}_{\phi} = \int d^4x \sqrt{-g}\  \left(-\frac{1}{2} \partial_\mu\phi \partial^\mu \phi\right) \,.
\label{inflation-c.1}
\end{equation}
It is found convenient to use the conformal time  $\zeta$ so that  
\begin{align}
\zeta &=  \int \frac{dt}{R(t)} \,.
\end{align}
Perturbations of the inflaton field should be considered along with the perturbations of the 
 gravitational field. 
The  general linear perturbations of the FRW metric can be parametrized as   
\begin{equation}
ds^2= R^2 \Big[ -(1+ 2 \alpha) d\zeta^2 
+ 2 \beta_i  dx^i d\zeta + (\delta_{ij} + h_{ij}) dx^i dx^j \Big] \,,
\label{inflation-c.18a}
\end{equation}
where $\alpha, \beta_i, h_{ij}$ parametrize the perturbations.  These perturbations can be 
 decomposed into scalar, vector and tensor components.  
 For $\beta_i$ we have 
 \begin{equation}
\beta_i= \nabla_i \beta+  \beta'_i \,, \  \
\nabla_i  \beta^{'i}=0 \,. 
\label{inflation-c.18b}
\end{equation}
where $\beta$ is the scalar and $\beta'_i$  the vector component.  We decompose  $h_{ij}$ 
so that 
\begin{align}
h_{ij} = 2 \gamma \delta_{ij} + 2 \nabla_i \nabla_j \kappa  + 2 \Delta_{(i}\ \kappa_{j)} +  \kappa_{ij} \,,\non
\nabla_i  \kappa^{ij} =0 \,, \  \kappa^{ij} \delta_{ij} =0 \,.  
\label{inflation-c.18d}
\end{align}
First we consider the scalar perturbations.
 It turns out that the scalar perturbation  of interest  is given by the combination
\begin{equation}
\sigma= R \left(\delta \phi -\gamma\frac {\phi'}{\cal H }\right) \,,
\label{inflation-c.18h}
\end{equation}
where the prime on $\phi$ stands for derivative of $\phi$ with respect to the conformal time $\zeta$
and ${\cal H}$ is the Hubble parameter in conformal time. 
Here $\sigma$ is a combination of  the scalar field and metric perturbations and is gauge invariant.  The quadratic part of action for $\sigma$ is given by
  \begin{equation}
  {\cal A}_{\sigma}\simeq \frac{1}{2} \int d\zeta d^3x \left[ \sigma^{'2} + \vec \nabla \sigma\cdot \vec\nabla \sigma +  \frac{z''}{z} \sigma^2\right] \,,
  \label{inflation-c.18i}
  \end{equation}  
  where $z=R \phi'/{\cal H}$. In the slow roll approximation one has $z''/z\simeq R''/R$. Thus 
    Eq.(\ref{inflation-c.18i}) takes the form   
    \begin{equation}
  {\cal A}_{\sigma}\simeq \frac{1}{2} \int d\zeta d^3x \left[ \sigma^{'2} + \vec \nabla \sigma\cdot \vec\nabla \sigma +  \frac{R''}{R} \sigma^2\right] \,.
  \label{inflation-c.18ai}
  \end{equation}  
  We now do a canonical quantization of the system by imposing the conformal time commutation relations  
\begin{gather}
\label{inflation-c.9a}
[\sigma(\zeta, \vec x) ,  \pi(\zeta, \vec x')]= i\hbar \delta(\vec x- \vec x')\,, \\
  [\sigma(\zeta, \vec x) \,, 
  \sigma(\zeta, \vec x')]
 = 0
  = [\pi(\zeta, \vec x) ,  \pi(\zeta, \vec x')] \,,
  \label{inflation-c.9}
 \end{gather}
where $\pi(\zeta, \vec x) = \sigma'$.  A Fourier expansion of $\sigma(\zeta, \vec x)$ leads to  
\begin{equation}
\sigma(\zeta, \vec x)= \int \frac{d^3k}{(2\pi)^{3/2}}  \left [ a(\vec k) \sigma(\vec k, \zeta) e^{i\vec k \cdot \vec x}
+ a^{\dagger}(\vec k) \sigma^*(\vec k, \zeta) e^{-i\vec k \cdot \vec x}\right] \,, 
\label{inflation-c.10}
\end{equation}
and quantization of the creation and the destruction operators $a(\vec k)$ and $a(\vec k)^{\dagger}$
so that 
\begin{gather}
 [a(\vec k), a^{\dagger}(\vec k')] = \delta (\vec k - \vec k')\,,\non
 ~ [a(\vec k), a(\vec k')] =0=   [a^\dagger(\vec k), a^{\dagger}(\vec k')]   \,. 
\label{inflation-c.11}
\end{gather}
where the Fourier component $\sigma(\vec k, \zeta)$  obey the equations of motion
\begin{equation}
 \sigma^{''}(\vec k, \zeta)+ (k^2- \frac{2}{\zeta^2})\sigma(\vec k, \zeta)=0 \,. 
\label{inflation-c.12}
\end{equation}
In Eq.(\ref{inflation-c.12}) we used $R''/R= 2/\zeta^2$. This result follows from the fact that for the case
when $H=\dot R/R$ is a constant the scale factor and the conformal time are related so that 
$R(\zeta) = - (H \zeta)^{-1}$. 
A specific solution to Eq.(\ref{inflation-c.12})~\cite{Bunch:1978yq,Chernikov:1968zm,Schomblond:1976xc} is  
 
\begin{equation}
 \sigma(\zeta,\vec k)= \sqrt{\frac{\hbar}{2 k}}\left(1- \frac{i}{k\zeta}\right) e^{-ik\zeta} \,. 
\label{inflation-c.14}
\end{equation}
Using the commutation relations of Eq.(\ref{inflation-c.11})
and the result above the  correlation function  $\langle~0|\sigma(\vec x)\sigma(\vec y)|0\rangle$ is given by 
\begin{equation}
\langle~0|\sigma(\vec x)\sigma(\vec y)|0\rangle= \int d^3k e^{i\vec k \cdot (\vec x-\vec y)} \frac{|\sigma(\zeta,\vec k)|^2}{(2\pi)^3 }\,. 
\label{inflation-c.15}
\end{equation}
The power spectrum for  $\sigma$ is given by 

\begin{equation}
{\cal P_\sigma}(k)= \frac{k^3}{2\pi^2}  {|\sigma(\zeta,\vec k)|^2}  \,. 
\label{inflation-c.17a}
\end{equation}
The  comoving curvature perturbation ${\cal R}$ is defined by 
\begin{align}
{\cal R}=  \frac{{\cal H}}{\phi'} \delta \phi - \gamma 
\label{inflation-c.17a}
\end{align}
Thus ${\cal R}= - \sigma/z$ snd  the  power spectrum for the curvature perturbation ${\cal P_R}(k)$ is given by   
\begin{equation}
{\cal P_R}(k)= \frac{k^3}{2\pi^2}  z^{-2} {|\sigma(\zeta,\vec k)|^2}  \,. 
\label{inflation-c.17}
\end{equation}
Now for the case when the wavelength is larger than the Hubble radius one has 
$k |\zeta| ~\ll~1$. Further using  the result $R(\zeta) = - (H \zeta)^{-1}$ and Eq.(\ref{inflation-c.14}) in 
Eq.(\ref{inflation-c.17})  and  reverting  to the 
cosmic time rather than using the conformal time one gets

\begin{align}
 {\cal P_R}  \simeq \frac{\hbar}{4 \pi^2} \left(H^4/ {\dot \phi}^2\right)_{k=RH}\,.
\end{align}

     Computation of the tensor power spectrum  is very similar to the scalar case except for the polarization and one gets
    \begin{align}
    {\cal P}_t= \frac{2\hbar}{\pi^2} \left(\frac{H}{M_{Pl}}\right)_{k=RH}^2\,.
    \end{align}
   ${\cal P_R}$ and ${\cal P}_t$ can be exhibited in terms of the inflaton potential. Here one finds 
   
   \begin{align}
   {\cal P_R}&= \frac{1}{12\pi^2}  \left(\frac{V^3}{M_{Pl}^6 V^{'2}}\right)_{k=RH}\,,\non   
   {\cal P}_t&= \frac{2}{3\pi^2} (\frac{V}{M_{Pl}^4})_{k=RH} \,. 
   \label{inflation-c.18q}
   \end{align} 
    Models of inflation are often parametrized in terms of the so called slow roll parameters $\epsilon, \eta$  defined by 
\begin{align}
\epsilon
&= \frac{1}{2 } \left(\frac{M_{Pl} V'}{V}\right)^2, ~~\eta = \left|\frac{M^2_{Pl}V''(\phi)}{V(\phi)}\right|\,.
   \label{slow-indices}
\end{align}
The spectral indices $n_s$ and $n_t$ (see Eqs.(\ref{ns}) and (\ref{nt})) 
are related to them by
\begin{align}
n_s= 1-6\epsilon + 2 \eta, ~~~n_t = -2 \epsilon, ~~ r = 16 \epsilon.
\label{ns-nt-r}
\end{align}
 where $r$ is defined by Eq.(\ref{ratio-r}).
Eq.(\ref{ns-nt-r}) is for the standard single-field inflation and  gives the prediction $n_t=-r/8$. Small deviations from this result  can occur for 
the effective single-field case and the size of the deviations will depend on the specifics of the inputs. 
A more detailed discussion of this is given in section 5.2.

\section{Appendix C:  Emergence of a flat inflation potential   for axions
  \label{appendixPotential}}

\begin{figure}
\centering
\includegraphics[width=0.5\textwidth]{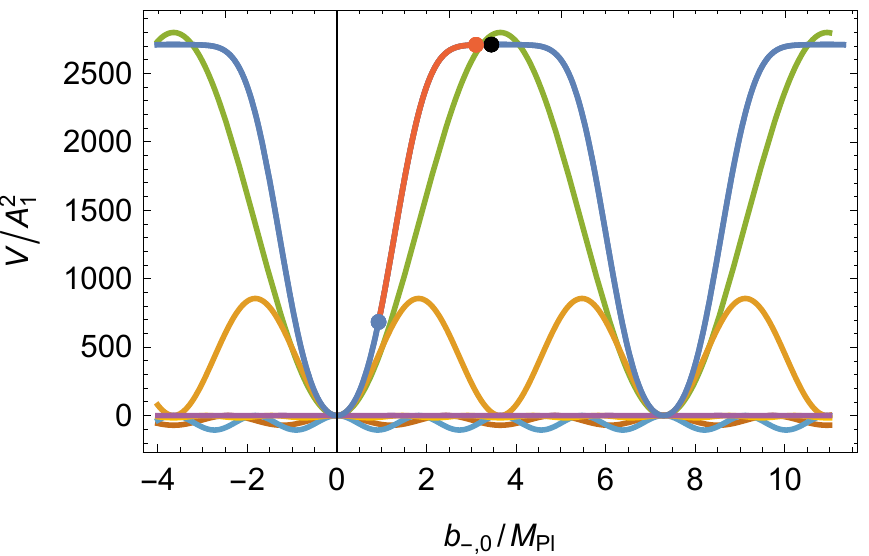}
\caption{The red and blue curve shows a generic example of a slow-roll potential in $q=3$ supergravity model. Here $A_2/A_1 = 8.9588\times 10^{-2}$, $A_3/A_1 = 4.1713\times 10^{-3}$, $\gamma_k = k$, $f/M_{Pl} = 0.8211$, $b_{-,0}/M_{Pl} = 3.448$, $N_\text{pivot} = 56.6$. For this case, the observables are $n_s = 0.9714$, $n_t = - 1.34 \times 10^{-5}$, $r = 1.07 \times 10^{-4}$. The black dot corresponds to the initial value of the field, the red dot to horizon exit, and the blue dot to the end of inflation. Other curves show contributions to the potential from individual cosine modes. 
Thus the green curve is plot of the first term in Eq.(\ref{c2})  in Appendix Sec.(\ref{appendixPotential})
while the brown curve is plot of the second  term in Eq.(\ref{c2}). The remaining terms in Eq.(\ref{c2}) 
 are curves with small amplitudes in the plot.
}
\label{fig7}
\end{figure}

\begin{figure}
\centering
\includegraphics[width=1\textwidth]{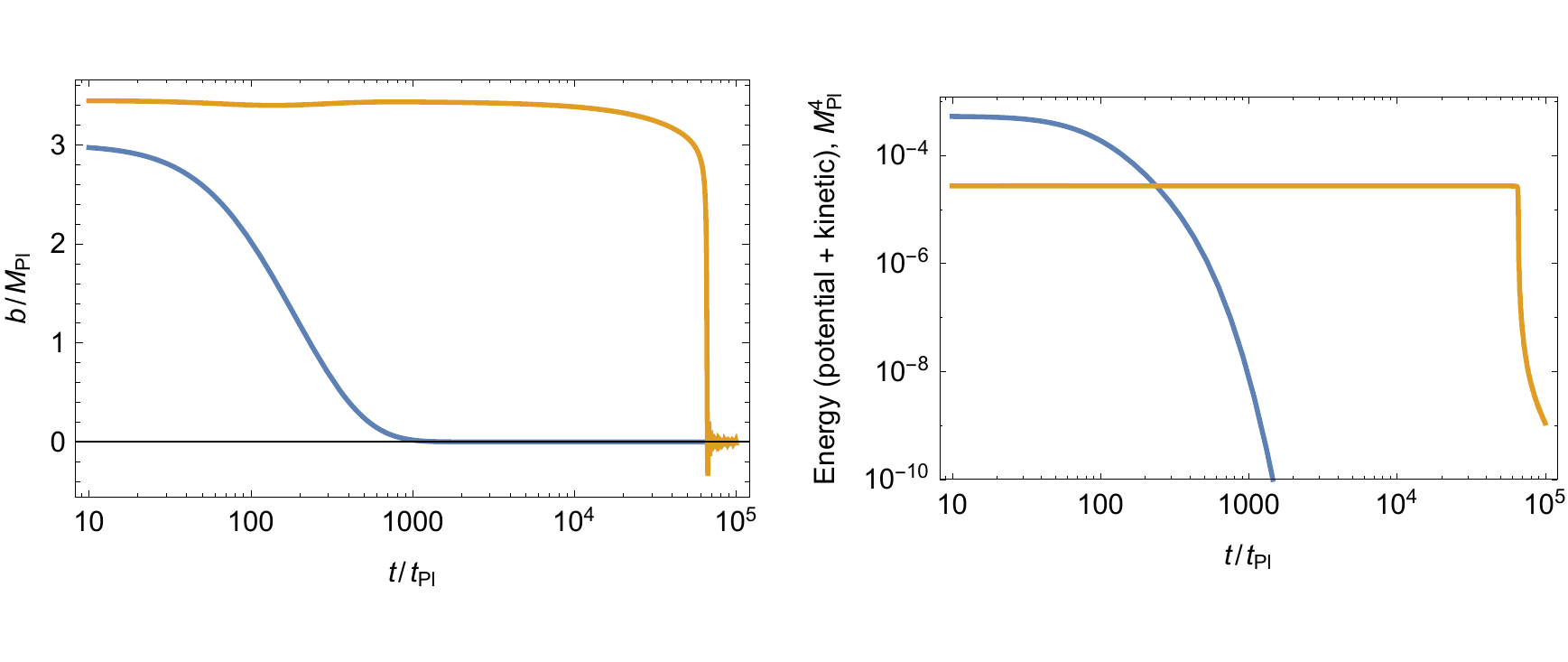}
\caption{Fast vs. slow field evolution for the case of Fig.(\ref{fig7}). Left panel: fast and slow field components as a function of time. Right panel: energy of the slow and fast field components as a function of time. Slow field energy is defined as ${E}_{\text{slow}} = {V}_{\text{slow}} \left( b_- \right) + \frac{1}{2} {\dot{b_-}}^{2}$. Fast field energy is defined as ${E}_{\text{fast}} = {V}_{\text{full}} \left( b_+, b_- \right) - {V}_{\text{slow}} \left( b_- \right) + \frac{1}{2} {\dot{b_+}}^{2}$. One can see that the fast component starts larger, but then the slow component overtakes it. Note the logarithmic scale of time. Here ${A}_{1} = {10}^{- 4} M_{Pl}^2$.}
\label{fig8}
\end{figure}

\begin{figure}
\includegraphics[width=1\textwidth]{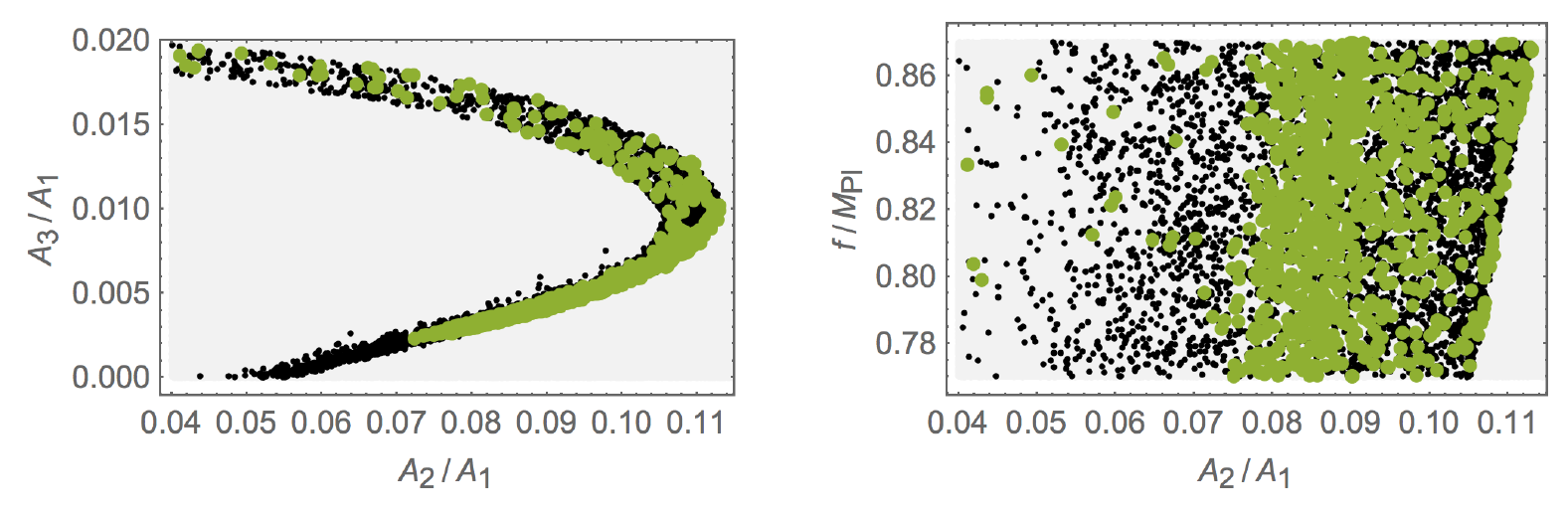}
\caption{ Left panel: $A_3/A_1$ vs. $A_2/A_1$ for the 
 experimentally acceptable points for the $q = 3$ supergravity model in the input parameters space
 for the case of one pair of axions.  Right panel: The same parameter space plotted in the $f/M_{Pl}$ 
 vs $A_2/A_1$.  The green and black parameter points have the same  meaning as in Fig.(\ref{fig3}).
Note the entire parameter space of this supergravity model exhibited here
has $f<M_{Pl}$. }
\label{fig9}
\end{figure}

\begin{figure}
\includegraphics[width=1\textwidth]{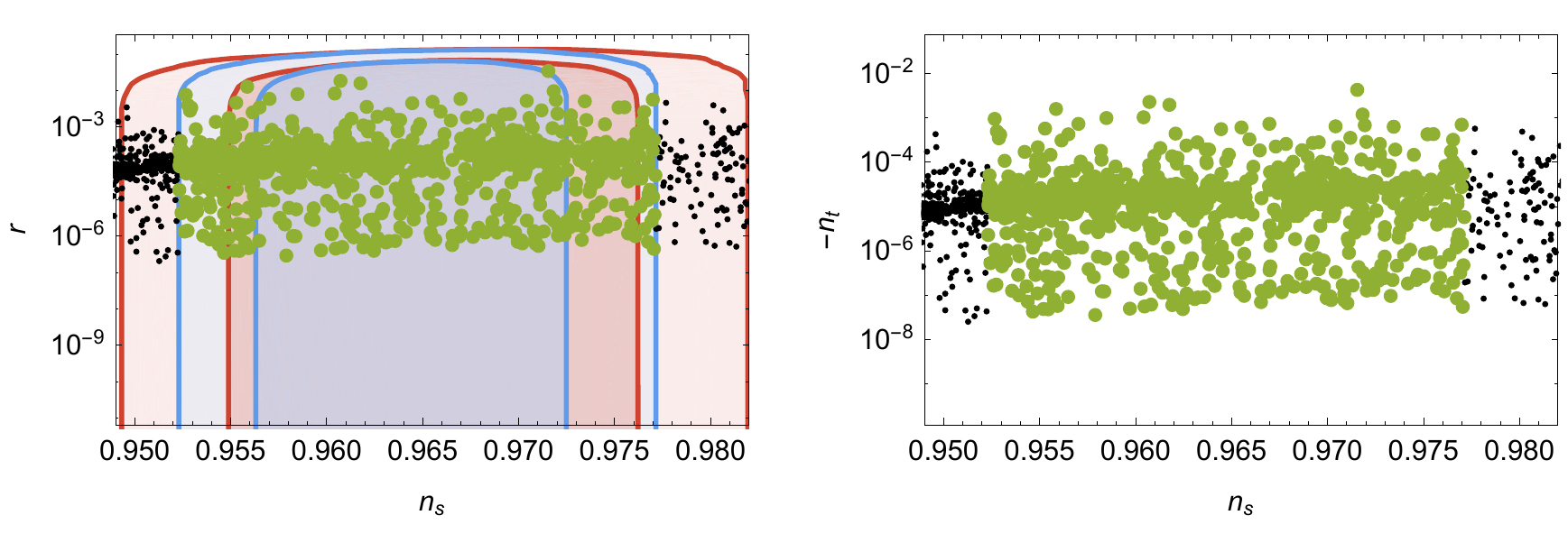}
\caption{Left panel:  Display of $r$ vs $n_s$ for the same parameter space as in Fig.(\ref{fig9}).
Right panel: Display $n_t$ vs $n_s$ for the same parameter space as in Fig.(\ref{fig9}).
The green and black parameter points have the same  meaning as in Fig.(\ref{fig3}).}
\label{fig10}
\end{figure}

Here we give an illustration of how a flat inflation potential for axions arises from a superposition of cosine functions 
with the specific case of 
 Fig.(\ref{fig7}) in mind.
 To exhibit this we begin by simplifying Eq.(\ref{sugrapot}) for the case of   Fig.(\ref{fig7}), i.e.,
 $q = 3$, and $\gamma_k = k$ for $k = 1, 2, 3$. Setting $M_{Pl} = 1$, we get:

\begin{align}
{V}_{3} \left( b \right) = &- \frac{1}{f^2} (4  {e}^{2  {f}^{2} + 2}  ( - 2  {A}_{1}^{2}  \left( 4  {f}^{4} + {f}^{2} + 1 \right) - e  {A}_{1}  \left( {A}_{2}  \left( 4  {f}^{2} + 3 \right)  {f}^{2} + 2  e  {A}_{3}  \left( 4  {f}^{4} + 5  {f}^{2} + 3 \right) \right) \non
&\qquad + {e}^{3}  {A}_{2}  {A}_{3}  \left( 4  {f}^{4} + 7  {f}^{2} + 12 \right) )) \left( 1 - \text{cos} \left( \frac{b}{\sqrt{2}  f} \right) \right) \non
&+ \frac{1}{f^2}(2  {e}^{2  {f}^{2} + 2}  ( {A}_{1}^{2}  \left( - \left( 4  {f}^{4} + {f}^{2} \right) \right) + 2  e  {A}_{1}  \left( {A}_{2}  \left( 8  {f}^{4} + 6  {f}^{2} + 4 \right) - e  {A}_{3}  \left( 4  {f}^{4} + 5  {f}^{2} + 6 \right) \right) \non
&\qquad + 4  {e}^{2}  {A}_{2}  \left( {A}_{2}  \left( 4  {f}^{4} + 5  {f}^{2} + 4 \right) + e  {A}_{3}  \left( 4  {f}^{4} + 7  {f}^{2} + 6 \right) \right) )) \left( 1 - \text{cos} \left( \frac{2 b}{\sqrt{2} f} \right) \right) \non
&+ \frac{1}{f^2}(4  {e}^{2  {f}^{2} + 3}  ( {A}_{1}  \left( 2  e  {A}_{3}  \left( 4  {f}^{4} + 5  {f}^{2} + 3 \right) - {A}_{2}  {f}^{2}  \left( 4  {f}^{2} + 3 \right) \right) \non
&\qquad + 2  {e}^{2}  {A}_{3}  \left( {A}_{2}  \left( 4  {f}^{4} + 7  {f}^{2} + 6 \right) + e  {A}_{3}  \left( 4  {f}^{4} + 9  {f}^{2} + 9 \right) \right) )) \left( 1 - \text{cos} \left( \frac{3  b}{\sqrt{2}  f} \right) \right) \non
&- 2  \left( {A}_{2}^{2} + 2  {A}_{1}  {A}_{3} \right)  {e}^{2  {f}^{2} + 4}  \left( 4  {f}^{2} + 5 \right)  \left( 1 - \text{cos} \left( \frac{4   b}{ \sqrt{2}f} \right) \right) \non
&- 4  {A}_{2}  {A}_{3}  {e}^{2  {f}^{2} + 5}  \left( 4  {f}^{2} + 7 \right)  \left( 1 - \text{cos} \left( \frac{5  b}{\sqrt{2}  f} \right) \right) \non
&- 2  {A}_{3}^{2}  {e}^{2  {f}^{2} + 6}  \left( 4  {f}^{2} + 9 \right)  \left( 1 - \text{cos} \left( \frac{6 b}{\sqrt{2} f} \right) \right)\,.
\label{c1}
\end{align}

Substituting $f = 0.8211$, $A_2/A_1 = 8.9588 \times 10^{-2}$, $A_3/A_1 = 4.1713 \times 10^{-3}$, as in Fig.(\ref{fig7}), we obtain:

\begin{align}
{V}_{3} \left( b \right)/A_1^2 &= 1398.96 \left( 1 - \text{cos} \left( \frac{b}{\sqrt{2}  f} \right) \right) + 427.466 \left( 1 - \text{cos} \left( \frac{2  b}{\sqrt{2} f} \right) \right) \non
& - 35.4939 \left( 1 - \text{cos} \left( \frac{3  b}{\sqrt{2}  f} \right) \right) - 52.9837 \left( 1 - \text{cos} \left( \frac{4 b}{\sqrt{2} f} \right) \right) \non
& - 8.28504 \left( 1 - \text{cos} \left( \frac{5  b}{\sqrt{2}  f} \right) \right) - 0.632442 \left( 1 - \text{cos} \left( \frac{6 b}{\sqrt{2} f} \right) \right)\,.
\label{c2}
\end{align}
Eq.(\ref{c2}) is the potential used in the analysis of Fig. (\ref{fig7}).

\clearpage

\clearpage
\begin{thebibliography}{99}

\bibitem{Guth:1980zm} 
  A.~H.~Guth,
  Phys.\ Rev.\ D {\bf 23}, 347 (1981).
  doi:10.1103/PhysRevD.23.347


\bibitem{Starobinsky:1980te} 
  A.~A.~Starobinsky,
  Phys.\ Lett.\  {\bf 91B}, 99 (1980).
  doi:10.1016/0370-2693(80)90670-X


\bibitem{Linde:1981mu} 
  A.~D.~Linde,
  Phys.\ Lett.\  {\bf 108B}, 389 (1982).
  doi:10.1016/0370-2693(82)91219-9


\bibitem{Albrecht:1982wi} 
  A.~Albrecht and P.~J.~Steinhardt,
  Phys.\ Rev.\ Lett.\  {\bf 48}, 1220 (1982).
  doi:10.1103/PhysRevLett.48.1220

\bibitem{Sato}
K.~Sato, 
Monthly Notices of the Royal Astronomical Society, Volume 195, Issue 3, 1 July 1981, Pages 467-479,https://doi.org/10.1093/mnras/195.3.467

\bibitem{Linde:1983gd} 
  A.~D.~Linde,
  Phys.\ Lett.\  {\bf 129B}, 177 (1983).
  doi:10.1016/0370-2693(83)90837-7

\bibitem{Linde:2005ht} 
  A.~D.~Linde,
  ``Particle physics and inflationary cosmology,''
  Contemp.\ Concepts Phys.\  {\bf 5}, 1 (1990)
  [hep-th/0503203].

\bibitem{Cheung:2007st} 
  C.~Cheung, P.~Creminelli, A.~L.~Fitzpatrick, J.~Kaplan and L.~Senatore,
  JHEP {\bf 0803}, 014 (2008)
  doi:10.1088/1126-6708/2008/03/014
  [arXiv:0709.0293 [hep-th]].


\bibitem{Mukhanov+}
V. F. Mukhanov and G. V. Chibisov, PisÕma Zh. Eksp. Teor. Fiz. 33, 549 (1981) [JETP Lett. 33, 532 (1981)]; S. W. Hawking, Phys. Lett. B 115, 295 (1982); A. A. Starobinsky, Phys. Lett. B 117, 175 (1982); A. H. Guth and S. Y. Pi, Phys. Rev. Lett. 49, 1110 (1982); J. M. Bardeen, P. J. Steinhardt, and M. S. Turner, Phys. Rev. D 28, 679 (1983).


\bibitem{Adam:2015rua} 
  R.~Adam {\it et al.} [Planck Collaboration],
  Astron.\ Astrophys.\  {\bf 594}, A1 (2016)
  doi:10.1051/0004-6361/201527101
  [arXiv:1502.01582 [astro-ph.CO]].


\bibitem{Ade:2015lrj} 
  P.~A.~R.~Ade {\it et al.} [Planck Collaboration],
  Astron.\ Astrophys.\  {\bf 594}, A20 (2016)
  doi:10.1051/0004-6361/201525898
  [arXiv:1502.02114 [astro-ph.CO]].



\bibitem{Array:2015xqh} 
  P.~A.~R.~Ade {\it et al.} [BICEP2 and Keck Array Collaborations],
  Phys.\ Rev.\ Lett.\  {\bf 116}, 031302 (2016)
  doi:10.1103/PhysRevLett.116.031302
  [arXiv:1510.09217 [astro-ph.CO]].



\bibitem{Freese:1990rb} 
  K.~Freese, J.~A.~Frieman and A.~V.~Olinto,
  Phys.\ Rev.\ Lett.\  {\bf 65}, 3233 (1990).
  doi:10.1103/PhysRevLett.65.3233


\bibitem{Adams:1992bn} 
  F.~C.~Adams, J.~R.~Bond, K.~Freese, J.~A.~Frieman and A.~V.~Olinto,
  Phys.\ Rev.\ D {\bf 47}, 426 (1993)
  doi:10.1103/PhysRevD.47.426
  [hep-ph/9207245].


\bibitem{Banks:2003sx} 
  T.~Banks, M.~Dine, P.~J.~Fox and E.~Gorbatov,
  JCAP {\bf 0306}, 001 (2003)
  doi:10.1088/1475-7516/2003/06/001
  [hep-th/0303252].

\bibitem{Svrcek:2006yi} 
  P.~Svrcek and E.~Witten,
  JHEP {\bf 0606}, 051 (2006)
  doi:10.1088/1126-6708/2006/06/051
  [hep-th/0605206].

\bibitem{Kim:2004rp} 
  J.~E.~Kim, H.~P.~Nilles and M.~Peloso,
  JCAP {\bf 0501}, 005 (2005)
  doi:10.1088/1475-7516/2005/01/005
  [hep-ph/0409138].


\bibitem{Long:2014dta} 
  C.~Long, L.~McAllister and P.~McGuirk,
  Phys.\ Rev.\ D {\bf 90}, 023501 (2014)
  doi:10.1103/PhysRevD.90.023501
  [arXiv:1404.7852 [hep-th]].


\bibitem{Berg:2009tg} 
  M.~Berg, E.~Pajer and S.~Sjors,
  Phys.\ Rev.\ D {\bf 81}, 103535 (2010)
  doi:10.1103/PhysRevD.81.103535
  [arXiv:0912.1341 [hep-th]].


\bibitem{Dimopoulos:2005ac} 
  S.~Dimopoulos, S.~Kachru, J.~McGreevy and J.~G.~Wacker,
  JCAP {\bf 0808}, 003 (2008)
  doi:10.1088/1475-7516/2008/08/003
  [hep-th/0507205].


\bibitem{Easther:2005zr} 
  R.~Easther and L.~McAllister,
  JCAP {\bf 0605}, 018 (2006)
  doi:10.1088/1475-7516/2006/05/018
  [hep-th/0512102].


\bibitem{Grimm:2007hs} 
  T.~W.~Grimm,
  Phys.\ Rev.\ D {\bf 77}, 126007 (2008)
  doi:10.1103/PhysRevD.77.126007
  [arXiv:0710.3883 [hep-th]].


\bibitem{Kallosh:2007cc} 
  R.~Kallosh, N.~Sivanandam and M.~Soroush,
  Phys.\ Rev.\ D {\bf 77}, 043501 (2008)
  doi:10.1103/PhysRevD.77.043501
  [arXiv:0710.3429 [hep-th]].


\bibitem{Olsson:2007he} 
  M.~E.~Olsson,
  JCAP {\bf 0704}, 019 (2007)
  doi:10.1088/1475-7516/2007/04/019
  [hep-th/0702109].


\bibitem{Battefeld:2007en} 
  D.~Battefeld and T.~Battefeld,
  JCAP {\bf 0705}, 012 (2007)
  doi:10.1088/1475-7516/2007/05/012
  [hep-th/0703012].


\bibitem{Kim:2011jea} 
  S.~A.~Kim, A.~R.~Liddle and D.~Seery,
  Phys.\ Rev.\ D {\bf 85}, 023532 (2012)
  doi:10.1103/PhysRevD.85.023532
  [arXiv:1108.2944 [astro-ph.CO]].


\bibitem{Kim:2010ud} 
  S.~A.~Kim, A.~R.~Liddle and D.~Seery,
  Phys.\ Rev.\ Lett.\  {\bf 105}, 181302 (2010)
  doi:10.1103/PhysRevLett.105.181302
  [arXiv:1005.4410 [astro-ph.CO]].

\bibitem{Rudelius:2014wla} 
  T.~Rudelius,
  JCAP {\bf 1504}, no. 04, 049 (2015)
  doi:10.1088/1475-7516/2015/04/049
  [arXiv:1409.5793 [hep-th]].


\bibitem{ArkaniHamed:2003mz} 
  N.~Arkani-Hamed, H.~C.~Cheng, P.~Creminelli and L.~Randall,
  JCAP {\bf 0307}, 003 (2003)
  doi:10.1088/1475-7516/2003/07/003
  [hep-th/0302034].


\bibitem{Kaplan:2003aj} 
  D.~E.~Kaplan and N.~J.~Weiner,
  JCAP {\bf 0402}, 005 (2004)
  doi:10.1088/1475-7516/2004/02/005
  [hep-ph/0302014].


\bibitem{Green:2009ds} 
  D.~Green, B.~Horn, L.~Senatore and E.~Silverstein,
  Phys.\ Rev.\ D {\bf 80}, 063533 (2009)
  doi:10.1103/PhysRevD.80.063533
  [arXiv:0902.1006 [hep-th]].

 
 \bibitem{Arvanitaki:2009fg} 
  A.~Arvanitaki, S.~Dimopoulos, S.~Dubovsky, N.~Kaloper and J.~March-Russell,
  Phys.\ Rev.\ D {\bf 81}, 123530 (2010)
  doi:10.1103/PhysRevD.81.123530
  [arXiv:0905.4720 [hep-th]].


\bibitem{Pajer:2013fsa} 
  E.~Pajer and M.~Peloso,
  Class.\ Quant.\ Grav.\  {\bf 30}, 214002 (2013)
  doi:10.1088/0264-9381/30/21/214002
  [arXiv:1305.3557 [hep-th]].

\bibitem{Marsh:2015xka} 
  D.~J.~E.~Marsh,
  Phys.\ Rept.\  {\bf 643}, 1 (2016)
  doi:10.1016/j.physrep.2016.06.005
  [arXiv:1510.07633 [astro-ph.CO]].


\bibitem{Nath:2016qzm}
  P.~Nath,
  ``Supersymmetry, Supergravity, and Unification,''
   Cambridge, Uk: Univ. Pr. (2016) 520 P. (Cambridge Monographs On Mathematical Physics).

\bibitem{Ellis:1982ws} 
  J.~R.~Ellis, D.~V.~Nanopoulos, K.~A.~Olive and K.~Tamvakis,
  Nucl.\ Phys.\ B {\bf 221}, 524 (1983).
  doi:10.1016/0550-3213(83)90592-8



 \bibitem{SUSYa}
  H.~Murayama, H.~Suzuki, T.~Yanagida and J.~Yokoyama,
  Phys.\ Rev.\ D {\bf 50}, 2356 (1994);
   E.~J.~Copeland, A.~R.~Liddle, D.~H.~Lyth, E.~D.~Stewart and D.~Wands,
  Phys.\ Rev.\ D {\bf 49}, 6410 (1994);
   E.~D.~Stewart,
  Phys.\ Rev.\ D {\bf 51}, 6847 (1995);
    G.~R.~Dvali, Q.~Shafi and R.~K.~Schaefer,
  Phys.\ Rev.\ Lett.\  {\bf 73}, 1886 (1994)
  doi:10.1103/PhysRevLett.73.1886
  [hep-ph/9406319];
 P.~Binetruy and G.~R.~Dvali,
  Phys.\ Lett.\ B {\bf 388}, 241 (1996);


\bibitem{Kors:2004hz} 
  B.~Kors and P.~Nath,
  Nucl.\ Phys.\ B {\bf 711}, 112 (2005).
  [hep-th/0411201].


\bibitem{SUSYb}
  M.~Yamaguchi,
  Class.\ Quant.\ Grav.\  {\bf 28}, 103001 (2011);
 M.~Kawasaki, M.~Yamaguchi and T.~Yanagida,
  Phys.\ Rev.\ Lett.\  {\bf 85}, 3572 (2000);
%
  M.~Yamaguchi and J.~'i.~Yokoyama,
  Phys.\ Rev.\ D {\bf 63}, 043506 (2001)
  [hep-ph/0007021];
%
D.~Baumann and D.~Green,
``Signatures of Supersymmetry from the Early Universe,''
Phys.\ Rev.\ D {\bf 85} (2012) 103520
[arXiv:1109.0292 [hep-th]].
  M.~Yamaguchi,
  Phys.\ Rev.\ D {\bf 64}, 063502 (2001)
  [hep-ph/0103045];
%
  M.~Kawasaki and M.~Yamaguchi,
  Phys.\ Rev.\ D {\bf 65}, 103518 (2002)
  [hep-ph/0112093];
%
  M.~Yamaguchi and J.~'i.~Yokoyama,
  Phys.\ Rev.\ D {\bf 68}, 123520 (2003)
  [hep-ph/0307373];
%
  P.~Brax and J.~Martin,
  Phys.\ Rev.\ D {\bf 72}, 023518 (2005)
  [hep-th/0504168];
%
  R.~Kallosh,
  Lect.\ Notes Phys.\  {\bf 738}, 119 (2008)
  [hep-th/0702059 [HEP-TH]];
%
  S.~C.~Davis and M.~Postma,
  JCAP {\bf 0803}, 015 (2008)
  [arXiv:0801.4696 [hep-ph]];
  F.~Takahashi,
  Phys.\ Lett.\ B {\bf 693}, 140 (2010)
  [arXiv:1006.2801 [hep-ph]];
  K.~Nakayama and F.~Takahashi,
  JCAP {\bf 1011}, 009 (2010)
  [arXiv:1008.2956 [hep-ph]];
%
  R.~Kallosh and A.~Linde,
  JCAP {\bf 1011}, 011 (2010)
  [arXiv:1008.3375 [hep-th]];
  M.~Kawasaki, N.~Kitajima and K.~Nakayama,
  Phys.\ Rev.\ D {\bf 83}, 123521 (2011)
  [arXiv:1104.1262 [hep-ph]];
  J.~Ellis, D.~V.~Nanopoulos and K.~A.~Olive,
  Phys.\ Rev.\ Lett.\  {\bf 111}, 111301 (2013)
  Erratum: [Phys.\ Rev.\ Lett.\  {\bf 111}, no. 12, 129902 (2013)]
  doi:10.1103/PhysRevLett.111.129902, 10.1103/PhysRevLett.111.111301
  [ JarXiv:1305.1247 [hep-th]];
   M.~Czerny, T.~Higaki and F.~Takahashi,
  JHEP {\bf 1405}, 144 (2014)
  doi:10.1007/JHEP05(2014)144
  [arXiv:1403.0410 [hep-ph]];
   T.~Kobayashi, S.~Uemura and J.~Yamamoto,
  Phys.\ Rev.\ D {\bf 96}, no. 2, 026007 (2017)
  doi:10.1103/PhysRevD.96.026007
  [arXiv:1705.04088 [hep-ph]];
  N.~Okada and Q.~Shafi,
  arXiv:1709.04610 [hep-ph].

\bibitem{BlancoPillado:2004ns} 
  J.~J.~Blanco-Pillado, C.~P.~Burgess, J.~M.~Cline, C.~Escoda, M.~Gomez-Reino, R.~Kallosh, A.~D.~Linde and F.~Quevedo,
  JHEP {\bf 0411}, 063 (2004)
  doi:10.1088/1126-6708/2004/11/063
  [hep-th/0406230].

\bibitem{Krippendorf:2009zza} 
  S.~Krippendorf and F.~Quevedo,
  JHEP {\bf 0911}, 039 (2009)
  doi:10.1088/1126-6708/2009/11/039
  [arXiv:0901.0683 [hep-th]].

\bibitem{Cicoli:2016olq} 
  M.~Cicoli, K.~Dutta, A.~Maharana and F.~Quevedo,
  JCAP {\bf 1608}, no. 08, 006 (2016)
  doi:10.1088/1475-7516/2016/08/006
  [arXiv:1604.08512 [hep-th]].



\bibitem{Peccei:1977ur} 
  R.~D.~Peccei and H.~R.~Quinn,
  Phys.\ Rev.\ D {\bf 16}, 1791 (1977).
  doi:10.1103/PhysRevD.16.1791


\bibitem{Weinberg:1977ma} 
  S.~Weinberg,
  Phys.\ Rev.\ Lett.\  {\bf 40}, 223 (1978).
  doi:10.1103/PhysRevLett.40.223


\bibitem{Wilczek:1977pj} 
  F.~Wilczek,
  Phys.\ Rev.\ Lett.\  {\bf 40}, 279 (1978).
  doi:10.1103/PhysRevLett.40.279


\bibitem{Halverson:2017deq} 
  J.~Halverson, C.~Long and P.~Nath,
  Phys.\ Rev.\ D {\bf 96}, no. 5, 056025 (2017)
  doi:10.1103/PhysRevD.96.056025
  [arXiv:1703.07779 [hep-ph]].

\bibitem{Cvetic:2009ez} 
  M.~Cvetic, J.~Halverson and R.~Richter,
  JHEP {\bf 1007}, 005 (2010)
  doi:10.1007/JHEP07(2010)005
  [arXiv:0909.4292 [hep-th]].


\bibitem{goldstone} 
 Y.~Nambu and G.~Jona-Lasinio,
  Phys.\ Rev.\  {\bf 122}, 345 (1961);
 J.~Goldstone,
  Nuovo Cim.\  {\bf 19}, 154 (1961);
  J.~Goldstone, A.~Salam and S.~Weinberg,
  Phys.\ Rev.\  {\bf 127}, 965 (1962).


\bibitem{Chamseddine:1982jx} 
  A.~H.~Chamseddine, R.~L.~Arnowitt and P.~Nath,
  Phys.\ Rev.\ Lett.\  {\bf 49}, 970 (1982).
  doi:10.1103/PhysRevLett.49.970


\bibitem{Cremmer:1982en} 
  E.~Cremmer, S.~Ferrara, L.~Girardello and A.~Van Proeyen,
  Nucl.\ Phys.\ B {\bf 212}, 413 (1983).
  doi:10.1016/0550-3213(83)90679-X

\bibitem{Nath:1983aw} 
  P.~Nath, R.~L.~Arnowitt and A.~H.~Chamseddine,
  Nucl.\ Phys.\ B {\bf 227}, 121 (1983).
  doi:10.1016/0550-3213(83)90145-1


\bibitem{Leach:2002ar} 
  S.~M.~Leach, A.~R.~Liddle, J.~Martin and D.~J.~Schwarz,
  Phys.\ Rev.\ D {\bf 66}, 023515 (2002)
  doi:10.1103/PhysRevD.66.023515
  [astro-ph/0202094].


\bibitem{Dias:2015rca} 
  M.~Dias, J.~Frazer and D.~Seery,
  JCAP {\bf 1512}, no. 12, 030 (2015)
  doi:10.1088/1475-7516/2015/12/030
  [arXiv:1502.03125 [astro-ph.CO]].


\bibitem{Lyth:1998xn} 
  D.~H.~Lyth and A.~Riotto,
  Phys.\ Rept.\  {\bf 314}, 1 (1999)
  doi:10.1016/S0370-1573(98)00128-8
  [hep-ph/9807278].

\bibitem{Liddle:2000cg} 
  A.~R.~Liddle and D.~H.~Lyth,
  Cambridge, UK: Univ. Pr. (2000) 400 p
\bibitem{Langlois:2010xc} 
  D.~Langlois,
  Lect.\ Notes Phys.\  {\bf 800}, 1 (2010)
  [arXiv:1001.5259 [astro-ph.CO]].
\bibitem{Baumann:2009ds} 
  D.~Baumann,
  arXiv:0907.5424 [hep-th].


\bibitem{Bunch:1978yq} 
  T.~S.~Bunch and P.~C.~W.~Davies,
  Proc.\ Roy.\ Soc.\ Lond.\ A {\bf 360}, 117 (1978).
  doi:10.1098/rspa.1978.0060


\bibitem{Chernikov:1968zm} 
  N.~A.~Chernikov and E.~A.~Tagirov,
  Ann.\ Inst.\ H.\ Poincare Phys.\ Theor.\ A {\bf 9}, 109 (1968).


\bibitem{Schomblond:1976xc} 
  C.~Schomblond and P.~Spindel,
  Ann.\ Inst.\ H.\ Poincare Phys.\ Theor.\  {\bf 25}, 67 (1976).

\bibitem{Arnowitt:1992aq} 
  R.~L.~Arnowitt and P.~Nath,
  Phys.\ Rev.\ Lett.\  {\bf 69}, 725 (1992).
  doi:10.1103/PhysRevLett.69.725

\bibitem{Lyth:1996im} 
  D.~H.~Lyth,
  Phys.\ Rev.\ Lett.\  {\bf 78}, 1861 (1997)
  doi:10.1103/PhysRevLett.78.1861
  [hep-ph/9606387].


\bibitem{Amin:2014eta} 
  M.~A.~Amin, M.~P.~Hertzberg, D.~I.~Kaiser and J.~Karouby,
  Int.\ J.\ Mod.\ Phys.\ D {\bf 24}, 1530003 (2014)
  doi:10.1142/S0218271815300037
  [arXiv:1410.3808 [hep-ph]].

 \bibitem{kilt}
 S.~Kachru, R.~Kallosh, A.~D.~Linde and S.~P.~Trivedi,
  Phys.\ Rev.\ D {\bf 68}, 046005 (2003). 

\bibitem{lvs}
   V.~Balasubramanian, P.~Berglund, J.~P.~Conlon and F.~Quevedo,
  JHEP {\bf 0503}, 007 (2005).

\bibitem{BlancoPillado:2004ns} 
  J.~J.~Blanco-Pillado, C.~P.~Burgess, J.~M.~Cline, C.~Escoda, M.~Gomez-Reino, R.~Kallosh, A.~D.~Linde and F.~Quevedo,
  ``{ Racetrack inflation},''
  JHEP {\bf 0411}, 063 (2004)
  doi:10.1088/1126-6708/2004/11/063
  [hep-th/0406230].
\bibitem{Lalak:2005hr} 
  Z.~Lalak, G.~G.~Ross and S.~Sarkar,
  ``{ Racetrack inflation} and assisted moduli stabilisation,''
  Nucl.\ Phys.\ B {\bf 766}, 1 (2007)
  doi:10.1016/j.nuclphysb.2006.06.041
  [hep-th/0503178].
\bibitem{Greene:2005rn} 
  B.~Greene and A.~Weltman,
  ``An Effect of alpha' corrections on {\bf racetrack inflation},''
  JHEP { 0603}, 035 (2006)
  doi:10.1088/1126-6708/2006/03/035
  [hep-th/0512135].
\bibitem{BlancoPillado:2006he} 
  J.~J.~Blanco-Pillado, C.~P.~Burgess, J.~M.~Cline, C.~Escoda, M.~Gomez-Reino, R.~Kallosh, A.~D.~Linde and F.~Quevedo,
  ``{Inflating in a better  racetrack},''
  JHEP {\bf 0609}, 002 (2006)
  doi:10.1088/1126-6708/2006/09/002
  [hep-th/0603129].
\bibitem{Sun:2006xv} 
  C.~Y.~Sun and D.~H.~Zhang,
  ``The Non-Gaussianity of { Racetrack Inflation} Models,''
  Commun.\ Theor.\ Phys.\  {\bf 48}, 189 (2007)
  doi:10.1088/0253-6102/48/1/038
  [astro-ph/0604298].
\bibitem{Brax:2007fe} 
  P.~Brax, A.~C.~Davis, S.~C.~Davis, R.~Jeannerot and M.~Postma,
  ``D-term Uplifted  {Racetrack Inflation},''
  JCAP {\bf 0801}, 008 (2008)
  doi:10.1088/1475-7516/2008/01/008
  [arXiv:0710.4876 [hep-th]].
\bibitem{Wen:2007ek} 
  W.~Y.~Wen,
  `{ Inflation in a refined racetrack},''
  arXiv:0712.0458 [hep-th].
\bibitem{Brax:2007fz} 
  P.~Brax, S.~C.~Davis and M.~Postma,
  ``The Robustness of n(s) <~ 0.95 in { racetrack inflation},''
  JCAP {\bf 0802}, 020 (2008)
  doi:10.1088/1475-7516/2008/02/020
  [arXiv:0712.0535 [hep-th]].
\bibitem{Wen:2008zz} 
  W.~Y.~Wen,
  ``Effects of open string moduli on { racetrack inflation},''
  Mod.\ Phys.\ Lett.\ A {\bf 23}, 1589 (2008).
  doi:10.1142/S0217732308027989
\bibitem{Brax:2008dn} 
  P.~Brax, C.~van de Bruck, A.~C.~Davis, S.~C.~Davis, R.~Jeannerot and M.~Postma,
  ``{ Racetrack Inflation} and Cosmic Strings,''
  JCAP {\bf 0807}, 018 (2008)
  doi:10.1088/1475-7516/2008/07/018
  [arXiv:0805.1171 [hep-th]].
\bibitem{Chen:2009nk} 
  H.~Y.~Chen, L.~Y.~Hung and G.~Shiu,
  ``{Inflation on an Open  Racetrack},''
  JHEP {\bf 0903}, 083 (2009)
  doi:10.1088/1126-6708/2009/03/083
  [arXiv:0901.0267 [hep-th]].
\bibitem{Badziak:2009eh} 
  M.~Badziak and M.~Olechowski,
  ``{ Inflation with racetrack} superpotential and matter field,''
  JCAP {\bf 1002}, 026 (2010)
  doi:10.1088/1475-7516/2010/02/026
  [arXiv:0911.1213 [hep-th]].
\bibitem{Allahverdi:2009rm} 
  R.~Allahverdi, B.~Dutta and K.~Sinha,
  ``Low-scale Inflation and Supersymmetry Breaking in { Racetrack Models},''
  Phys.\ Rev.\ D {\bf 81}, 083538 (2010)
  doi:10.1103/PhysRevD.81.083538
  [arXiv:0912.2324 [hep-th]].
\bibitem{Olechowski:2010zz} 
  M.~Olechowski,
  ``{Inflation with racetrack} superpotential and matter field,''
  J.\ Phys.\ Conf.\ Ser.\  {\bf 259}, 012028 (2010).
  doi:10.1088/1742-6596/259/1/012028
\bibitem{Badziak:2010qy} 
  M.~Badziak,
  ``F-term uplifted { racetrack inflation},''
  arXiv:1005.5537 [hep-th].


\bibitem{Higaki:2014pja} 
  T.~Higaki and F.~Takahashi,
  JHEP {\bf 1407}, 074 (2014)
  doi:10.1007/JHEP07(2014)074
  [arXiv:1404.6923 [hep-th]].

\bibitem{Higaki:2014mwa} 
  T.~Higaki and F.~Takahashi,
  Phys.\ Lett.\ B {\bf 744}, 153 (2015)
  doi:10.1016/j.physletb.2015.03.052
  [arXiv:1409.8409 [hep-ph]].

\bibitem{Kadota:2016jlw} 
  K.~Kadota, T.~Kobayashi, A.~Oikawa, N.~Omoto, H.~Otsuka and T.~H.~Tatsuishi,
  JCAP {\bf 1610}, no. 10, 013 (2016)
  doi:10.1088/1475-7516/2016/10/013
  [arXiv:1606.03219 [hep-ph]].

\bibitem{Kobayashi:2016vcx} 
  T.~Kobayashi, A.~Oikawa, N.~Omoto, H.~Otsuka and I.~Saga,
  Phys.\ Rev.\ D {\bf 95}, no. 6, 063514 (2017)
  doi:10.1103/PhysRevD.95.063514
  [arXiv:1609.05624 [hep-ph]].

\bibitem{Ernst:2018bib} 
  A.~Ernst, A.~Ringwald and C.~Tamarit,
  arXiv:1801.04906 [hep-ph].


\end{thebibliography}
\end{document}